\documentclass[letterpaper,11pt]{article}

\usepackage{fancyhdr}
\usepackage{amsmath}
\usepackage{amsbsy}
\usepackage{amssymb}
\usepackage{amsthm}
\usepackage{amscd}
\usepackage{amsfonts}
\usepackage{supertabular}
\usepackage{graphics}
\usepackage{graphicx}
\usepackage{verbatim}
\usepackage[title]{appendix}
\usepackage{epsfig}
\usepackage{xspace}
\usepackage{euscript}
\usepackage{alltt}
\usepackage{times}
\usepackage[square,numbers,sort&compress]{natbib}
\usepackage[colorlinks]{hyperref}
\usepackage{setspace}

\usepackage{authblk}
\usepackage{color,soul}
\usepackage{tikz}
\usetikzlibrary{positioning, shapes.geometric, arrows.meta} 
\usepackage{svg}
\usepackage{graphicx} 

\usepackage{nccmath}
\usepackage{lscape}

\usepackage[colorinlistoftodos]{todonotes}

\usepackage{algorithm}
\usepackage{algorithmic}

\usepackage{multirow}

\begin{document}

	\title{Odd Elasticity in Strain-Gradient Continua: \\ Anisotropy Classification and Non-Hermitian Elastodynamics
	}

	\author{Payam Poorsolhjouy\thanks{E-mail address: p.poorsolhjouy@tue.nl}}
	\affil{Applied Mechanics Group, Department of the Built Environment \\ Eindhoven University of Technology, De Groene Loper 6, 5612 AP Eindhoven, The Netherlands}

	\maketitle
	
	\begin{abstract}
		Odd elasticity extends continuum mechanics to active nonconservative systems containing internal energy sources, such as robotic active metamaterials with nonreciprocal feedback-controlled elements. Modeling these systems as effective continua requires relaxing Maxwell-Betti reciprocity, which results in stiffness tensors lacking major symmetry. In the present contribution, we extend this framework to three-dimensional strain-gradient odd elasticity to incorporate the length scales inherent in these heterogeneous systems. By applying the full hierarchy of 3D point-group symmetries along chiral and achiral branches, we derive the exhaustive set of fourth-, fifth-, and sixth-order odd elastic stiffness tensors. Further, utilizing the dynamic principle of virtual work, we formulate the equilibrium equations, higher-order boundary conditions, and the generalized acoustic tensor for elastodynamic analysis of 3D strain-gradient odd elastic systems. 
		
		The results demonstrate that nonreciprocal moduli survive even under full isotropy in the sixth-order stiffness, under hemitropy in the fifth-order stiffness, and under cubic symmetry in fourth-order stiffness. In contrast, the elastodynamic analysis reveals that under isotropic and hemitropic limits, despite the underlying odd stiffness moduli, the acoustic tensor remains unconditionally Hermitian. Non-Hermitian elastodynamics is shown to require anisotropy. By numerically analyzing a transversely hemitropic odd elastic continuum, we demonstrate that this symmetry reduction results in non-Hermitian elastodynamic phenomena including exceptional points in the dispersion spectra and active energy exchange through simultaneous amplification and attenuation of the elastic wave components. These results provide the necessary theoretical foundation for future multiscale homogenization frameworks for anisotropic strain-gradient active microstructures, and for the design of nonreciprocal strain-gradient metamaterials.
	\end{abstract}

	\section*{Keywords}

	Odd elasticity, Strain-gradient elasticity, Non-Hermitian elastodynamics, Anisotropy, Elastic symmetry groups, Exceptional points

	\section{Introduction}\label{sec:intro}
	
	The elastic mechanical behavior of passive materials is classically described by Green elasticity, also known as hyperelasticity \cite{truesdell1952mechanical}. Within the bounds of elastic behavior (and therefore in the absence of irreversible dissipative phenomena such as damage and plasticity), these materials conserve energy and cannot perform net mechanical work on their surroundings over closed deformation cycles. Consequently, their constitutive modeling relies on the assumption of the existence of a strain energy potential. In recent years, however, there has been growing interest in modeling \textit{active nonconservative systems}, whose microscopic constituents contain \textit{energy sources}, as \textit{effective continua}. Since these energy sources enable such systems to perform net mechanical work over closed deformation cycles, the existence of a strain energy potential cannot be assumed. As a consequence, their constitutive relationships need to become non-potential, thereby violating Maxwell-Betti reciprocity.
	
	This nonconservative framework, recently formalized as \textit{odd elasticity} \cite{scheibner2020odd, fruchart2023odd}, allows active nonreciprocal microscopic interactions to break the major symmetry of stiffness tensors and manifest as antisymmetric (odd) moduli. By explicitly capturing these nonreciprocal active interactions, odd elasticity has emerged as a widely used framework for the design and analysis of engineered robotic metamaterials equipped with distributed sensors and actuators \cite{brandenbourger2019non, chen2021realization, zheng2025metamaterial, cheng2026elastodynamics, dauchot2026active, veenstra2025adaptive, binysh2026more} and for modeling the complex behavior of active biological matter (e.g., muscle fibers, cell monolayers), active fluids, and topological defects \cite{shankar2024active, santhosh2026achiral, lee2026odd, ishimoto2023odd, banerjee2026emergent, braverman2021topological}. The odd moduli in these systems are directly governed by the programmable gain of nonreciprocal feedback loops in engineered robotic metamaterials \cite{brandenbourger2019non, chen2021realization}, or by the activity level of energy-consuming motors in biological systems \cite{shankar2024active, ishimoto2023odd}. Beyond their effects on the static constitutive relations, these nonreciprocal interactions have also been shown to result in non-Hermitian elastodynamics. While non-Hermitian phenomena—such as active wave amplification, directional nonreciprocity, and exceptional points (EPs)—can be engineered using time-varying properties \cite{nassar2017modulated,nassar2020nonreciprocity}, complex moduli with active gain and loss \cite{rosa2021exceptional, duan2025exceptional}, or non-local feedback controls \cite{rosa2020dynamics}, odd elasticity provides a distinct pathway via nonconservative nonreciprocal interactions \cite{ghatak2020observation, scheibner2020non, christensen2024perspective, zhou2026curved}.
	
	Unlike earlier phenomenological approaches for active biological systems (such as cytoskeletons, active polar gels, and self-organizing tissues \cite{kruse2004asters, kruse2005generic, juelicher2007active, joanny2009active, sasai2013cytosystems, needleman2017active, caldarelli2024self}), which typically superimpose an autonomous, strain-independent ``active stress'' onto a classical conservative continuum \cite{ramaswamy2010mechanics, marchetti2013hydrodynamics}, odd elasticity explicitly models systems where the microstructure uses internal energy sources to respond nonreciprocally to local deformations. By relaxing Maxwell-Betti reciprocity, in odd elasticity, the system's activity is embedded directly into the material's constitutive relations. This active, nonconservative breaking of major symmetry must also be distinguished from asymmetries that may arise in passive conservative media (e.g., asymmetric tangent stiffness tensors known to emerge as a geometric consequence of mapping Cauchy stress to the displacement gradient under finite deformations and prestresses \cite{yin2023anisotropy}, or the breakage of minor symmetry in chiral systems with micro-rotations \cite{yilmaz2026macroscale}). As noted by Yavari et al. \cite{yavari2025universal, yavari2025nonlinear}, odd elasticity, in its first-gradient form, finds its historical precedent in nonpotential linear Cauchy elasticity \cite{cauchy1828equations}, which was classically deemed inapplicable to passive materials because it permits the extraction of net work over closed deformation cycles \cite{coleman1962mechanical}. While this non-potential behavior is a physical requirement for active systems, as noted in \cite{yavari2025nonlinear, ostoja2024generalizing, wolfgram2026scaling}, the theory must still satisfy objectivity and the balance of angular momentum. Accordingly, a rigorous formulation of odd elasticity, though it involves breaking major symmetry, has to retain minor symmetries, as explicitly enforced in recent odd elasticity contributions \cite{lee2026odd}.
	
	Although initially explored mainly in the context of active chiral materials—where broken reflection symmetry drives nonreciprocal shear-normal couplings \cite{scheibner2020odd, fruchart2020symmetries, lee2026odd}—recent studies have expanded odd elasticity to achiral anisotropic active systems, such as muscle fibers \cite{shankar2024active} and active nematic cell monolayers \cite{santhosh2026achiral}. Concurrently, odd elasticity has been extended beyond the classical Cauchy continuum by introducing active micro-rotational degrees of freedom in odd Cosserat and micropolar theories \cite{chen2021realization, surowka2023odd}, or through reduced-dimensional structural elements like active plates and shells \cite{wang2024non, lai2024odd, lai2024anomalous, fossati2024odd, wolfgram2025odd}. These frameworks have also been utilized to demonstrate the aforementioned non-Hermitian dynamic phenomena in first-gradient Cauchy continua \cite{scheibner2020odd, fruchart2020symmetries, gao2022non}, micropolar media \cite{chen2021realization, surowka2023odd}, and reduced-dimensional plates \cite{wang2024non, fossati2024odd}. However, a comprehensive 3D bulk theory to capture the dispersive effects of internal length scales and explore non-Hermitian elastodynamics via higher gradients of displacement remains absent.
	
	Microstructural heterogeneities inherent in metamaterials, such as spatially varying hinge rotations or compliant pantographic junctions \cite{hild2021multiscale, durand2022predictive, czajkowski2022conformal}, naturally give rise to higher-gradient continuum effects. To address these, strain-gradient theories, which enrich the kinematics with higher gradients of displacement \cite{mindlin1963microstructure, mindlin1965second, mindlin1968first, triantafyllidis1986gradient, toupin1962elastic, Aif1992, Fle1997, Ask2002, aifantis2011gradient, askes2011gradient, reiher2018finite, faghidian2020higher, weeger2021numerical, molavitabrizi2023second}, have been widely developed and used for capturing high-frequency dispersive phenomena \cite{Sui2001c, dontsov2013physical, madeo2015wave} and for connecting microstructural features to macroscopic wave behavior \cite{Sui2001b, Payam-BandGap, PayamJMPS, misra2019longitudinal, nejadsadeghi2020role}. Accordingly, to properly incorporate the length-scale effects in nonconservative metamaterials, which are made by augmenting such heterogeneous microstructures with active, nonreciprocal feedback loops (e.g., \cite{chen2021realization,brandenbourger2019non,scheibner2020odd}), odd elasticity needs to be extended to the strain-gradient framework. Furthermore, the effects of the directional distribution of these heterogeneous microstructural features (both active and passive) on the configuration of independent components in higher-order odd elastic stiffness tensors need to be understood. For conservative materials, the anisotropy classification of higher-gradient stiffness tensors \cite{mousavi2016analysis, tahaei2017size, lazar2015non, auffray2013matrix, auffray2019complete, quang2021classification}, and their links to microstructural orientation distributions \cite{Payam-Gibbs, pirmoradi2024anisotropic}, have been extensively investigated. On the other hand, in the first-gradient domain, the effects of anisotropy on major-asymmetric stiffness tensors have been extensively studied (see \cite{ostrosablin2017symmetry, yavari2025nonlinear} for 3D and \cite{wolfgram2026scaling} for 2D).
	Recently, generalized tensorial representations based on director tensors have also been developed to accommodate anisotropic odd elastic couplings within classical first-gradient continua \cite{wolfgram2026generalization}. Nevertheless, a classification of the effects of different levels of anisotropy on the fourth-, fifth-, and sixth-order stiffness tensors in 3D strain-gradient odd elasticity has yet to be established.
	
	In the present contribution, we establish the mechanical foundation for odd elasticity in anisotropic 3D strain-gradient continua and apply it to study elastodynamics and wave propagation. In Section \ref{sec:symmetries}, by revoking Maxwell-Betti reciprocity in Mindlin's Form II second-gradient framework \cite{mindlin1965second}, we formulate the constitutive relationships for strain-gradient odd elasticity. Subsequently, by applying the full hierarchy of three-dimensional point-group symmetries along independent chiral and achiral branches, we systematically derive the fourth-, fifth-, and sixth-order odd elastic stiffness tensors for materials with different levels of anisotropy. We demonstrate that in second-gradient odd elasticity, even in the completely isotropic case, nonreciprocal stiffness moduli can survive. 
	
	Subsequently, in Section \ref{sec:waves}, utilizing the dynamic principle of virtual work, we derive the governing equilibrium equations, higher-order boundary conditions, equations of motion, and the generalized acoustic tensor for non-Hermitian elastodynamic analysis of strain-gradient odd elastic systems. We then apply this framework to solve the elastic wave propagation problem for specific anisotropy classes. The results presented in Section \ref{sec:results} show that in highly symmetric media (isotropic and chiral hemitropic), the acoustic tensor remains Hermitian, despite the existence of odd elastic moduli. However, in higher levels of anisotropy, the acoustic tensor can become non-Hermitian, resulting in the emergence of EPs and active energy exchange. Numerical elastodynamic studies are performed on physically realistic stiffness and length scales extracted from a previous study of a conservative granular assembly \cite{misra2017elastic}, upon which we have superimposed odd-elastic perturbations. Concluding remarks follow in Section \ref{sec:conclusion}.

	\section{Anisotropic Constitutive Relations in Second-Gradient Odd Elasticity}
	\label{sec:symmetries}
	
	The most general form of constitutive relationships for second-gradient elasticity is given as
	\begin{equation}
		\begin{array}{lcl}
			\tau_{ij} & = & C_{{(ij)}\;{(kl)}}\epsilon_{kl} + F_{{(ij)}\;k{(lm)}}\kappa_{klm} \, , \\
			\mu_{ijk} & = & G_{i{(jk)}\;{(lm)}}\epsilon_{lm} + A_{{i(jk)}\;{l(mn)}}\kappa_{lmn} \, ,
		\end{array}
		\label{Const-Macro}
	\end{equation}
	where $\mathbb{C}$ and $\mathbb{A}$ represent the fourth- and sixth-order stiffness tensors corresponding to the first and second gradient elasticity, while $\mathbb{F}$ and $\mathbb{G}$ are the fifth-order coupling elasticity tensors. Throughout the paper, the summation convention over repeated indices is implied. It is noted that both stress and strain measures are symmetric, $\tau_{ij}=\tau_{ji}$, $\epsilon_{ij}=\epsilon_{ji}$, $\mu_{ijk}=\mu_{ikj}$, and $\kappa_{ijk}=\kappa_{ikj}$ \cite{auffray2019complete,pirmoradi2024anisotropic,yavari2025nonlinear}. Accordingly, all stiffness tensors have both minor symmetries, which are shown by the parentheses implemented in the indices in Eq.~\eqref{Const-Macro}.
	We emphasize that in classical hyper-elasticity, the existence of a strain energy density potential enforces major symmetry on even-order constitutive tensors (i.e., $C_{ijkl} = C_{klij}$ and $A_{ijklmn} = A_{lmnijk}$) and dictates that the odd-order coupling tensors are transposes of one another (i.e., $G_{ijklm} = F_{lmijk}$). However, here and in the absence of an energy potential, Maxwell-Betti reciprocity is relaxed. Accordingly, major symmetry of the even-order stiffness tensors is lost and the two coupling tensors, $\mathbb{F}$ and $\mathbb{G}$, are independent.
	
	The constitutive relationships given in Eq.~\eqref{Const-Macro} can be reformulated using Voigt notation, by writing stress and kinematic measures in 1D array form and the stiffness tensors in matrix form as
	\begin{equation} 
		\begin{array}{lcllll}
			\vspace{2mm}
			\tau_\alpha & = & C_{\alpha\beta}\epsilon_\beta+F_{\alpha\gamma}\kappa_\gamma, & \qquad  \alpha: \text{1 to 6},  & \quad   \beta: \text{1 to 6}, & \quad \gamma: \text{1 to 18}, \  \\ 
			\mu_\alpha & = & G_{\alpha\beta}\epsilon_\beta+A_{\alpha\gamma}\kappa_\gamma,  & \qquad  \alpha: \text{1 to 18},  & \quad   \beta: \text{1 to 6}, & \quad \gamma: \text{1 to 18} \,.
		\end{array}
		\label{Const-Macro-Voigt}
	\end{equation}
	The explicit correspondence between the indices of the second- and third-order stress and strain tensors and their corresponding Voigt representations ($\alpha, \beta, \gamma$) is detailed in Appendix \ref{app:voigt} (Tables \ref{ij-alpha} and \ref{ijk-alpha}).
	To determine the odd elasticity tensors for any level of anisotropy, we enforce the principle of material symmetry. For each level of anisotropy, the constitutive tensors must remain invariant under any orthogonal transformation $\mathbf{Q}$ belonging to the material's symmetry point group:
	\begin{equation}
		\begin{array}{lcl}
			\vspace*{1mm}
			{C}_{pqrs} & = & C_{ijkl} Q_{ip}Q_{jq}Q_{kr}Q_{ls}  \, , \\
			\vspace*{1mm}
			{F}_{pqrst} & = & F_{ijklm} Q_{ip}Q_{jq}Q_{kr}Q_{ls}Q_{mt} \, , \\
			\vspace*{1mm}
			{G}_{pqrst} & = & G_{ijklm} Q_{ip}Q_{jq}Q_{kr}Q_{ls}Q_{mt} \, , \\
			{A}_{pqrstu} & = & A_{ijklmn} Q_{ip}Q_{jq}Q_{kr}Q_{ls}Q_{mt}Q_{nu} \, .
		\end{array}
		\label{CFGA-trans}
	\end{equation}
	
\begin{figure}[h]
		\centering
		\includegraphics[scale=0.7]{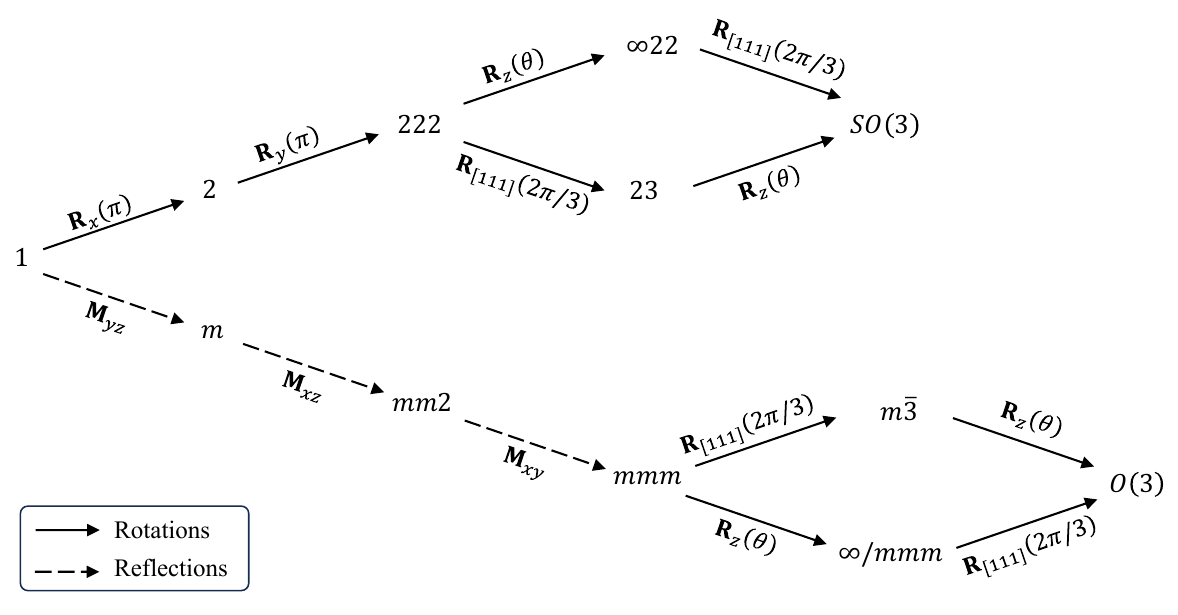}
		\caption{Hierarchy of elastic symmetry classes. The upper and lower branches contain chiral and achiral symmetry groups, respectively. Solid and dashed arrows denote rotations and reflections, respectively.}
		\label{fig:symmetry_hierarchy}
	\end{figure}

	To systematically evaluate the effects of these transformations, we organize the symmetry point groups into a hierarchical structure (Figure \ref{fig:symmetry_hierarchy}) containing two distinct branches: the \textit{chiral branch} consisting exclusively of proper rotations (orientation-preserving transformations), and the \textit{achiral branch} incorporating improper rotations (mirror reflections and central inversion). 
	
	The following subsections detail the derivation of odd elastic strain-gradient stiffness tensors for all the representative symmetry classes outlined in Figure \ref{fig:symmetry_hierarchy}. The explicit matrix representations for all discussed anisotropy levels, including identities between the different components, are provided in Appendix \ref{app:matrices}. Furthermore, a comprehensive classification detailing the number of independent components across all fundamental crystallographic and continuous point groups (including, in addition to the groups studied in the main text, the tetragonal, trigonal, and hexagonal families) is presented in the Appendix section \ref{app:expanded_classification} (Table \ref{tab:Expanded_Summary}).

	\subsection{The Chiral Branch (Proper Rotations)}
	In this branch, the material microstructure lacks any plane of reflection or center of inversion, and therefore all symmetry groups consist only of orientation-preserving rotation transformations, i.e., $\mathbf{Q_+}$ with $\det(\mathbf{Q_+}) = 1$. Accordingly, all materials in this group are noncentrosymmetric and chiral. 
	
	We depart from the case of a completely anisotropic (triclinic) material (point group $1$) as the most general chiral anisotropic material that contains no rotation or reflection symmetry. For triclinic materials, the elasticity tensors are all fully populated. Written in Voigt notation and considering the minor symmetries of the stress and strain measures, $\mathbb{C}$ can be written as a $6\times6$ matrix with 36 independent components. Similarly, $\mathbb{A}$ can be written as an $18\times18$ matrix with 324 independent components, and $\mathbb{F}$ and $\mathbb{G}$ are written as $6\times18$ and $18\times6$ matrices with 108 independent components each. 
	
	\subsubsection{Chiral Monoclinic (Point Group 2)}
	\label{sec:chiralmono}
	This class possesses a single axis of two-fold (i.e., $180^\circ$) rotation symmetry. Without loss of generality, we assume the \textit{x}-axis to be the axis of rotation symmetry. The transformation tensor for this rotation is given as
	\begin{equation}
		\mathbf{R}_x(\pi) = \begin{bmatrix} 1 & 0 & 0 \\ 0 & -1 & 0 \\ 0 & 0 & -1 \end{bmatrix},
		\label{pathA-Rot-x-180}
	\end{equation}
	which leaves $x$ unchanged and maps $y \to -y$ and $z \to -z$. We define $N_i$ as the number of times index $i$ appears in the subscript of a given tensor component, and $p_i \equiv N_i \pmod 2$ as its parity ($p_i \in \{0,1\}$). 
	Substituting Eq.~\eqref{pathA-Rot-x-180} into Eq.~\eqref{CFGA-trans}, any tensor component on the right hand side is multiplied by $(-1)^{N_2 + N_3}$. For the component to remain invariant, $N_2 + N_3$ must be even, which means
	\begin{equation}
		\begin{aligned}
			C_{ijkl}, \, A_{ijklmn} = \;
			\begin{cases}
				\text{nonzero} & \text{if index 1 appears even number of times ($p_1=0$),} \\
				\text{zero} & \text{if index 1 appears odd number of times ($p_1=1$), } 
			\end{cases}
		\end{aligned}
		\label{pathA-Mono-CA}
	\end{equation}
	and
	\begin{equation}
		\begin{aligned}
			F_{ijklm}, \, G_{ijklm} = \;
			\begin{cases}
				\text{nonzero} & \text{if index 1 appears odd number of times ($p_1=1$),} \\
				\text{zero} & \text{if index 1 appears even number of times  ($p_1=0$). } 
			\end{cases}
		\end{aligned}
		\label{pathA-Mono-FG}
	\end{equation}
	
	To identify the nonzero tensor components, as elaborated in Appendix section \ref{app:sec:mono}, Eq.~\eqref{parity-tensor}, the second- and third-order kinematic and stress measures are decomposed into subsets $\mathcal{V}_0$ and $\mathcal{K}_0$ (containing components where index 1 appears an even number of times, $p_1=0$), and $\mathcal{V}_1$ and $\mathcal{K}_1$ (containing components where index 1 appears an odd number of times, $p_1=1$). 
	For even-order tensors ($\mathbb{C}$ and $ \mathbb{A}$), the condition $p_1=0$ dictates that the parity of index 1 must be identical in the stress and strain measures. Conversely, for odd-order tensors ($\mathbb{F}$ and $\mathbb{G}$), the condition $p_1=1$ dictates that the parity of index 1 in the stress and strain must be unequal. 
	
	Accordingly, following Eqs.~\eqref{parity-tensor}, \eqref{parity-voigt}, the fourth-, fifth-, and sixth-order elasticity tensors in monoclinic materials contain $20$, $52$, and $164$ independent components, each. The matrix representation of the fourth- and sixth-order odd elastic stiffness for chiral monoclinic materials are given in Eqs.~\eqref{Cij-mono} and \eqref{A-mono}. Similarly, the fifth-order stiffness tensor, $\mathbb{F}$, is given in Eq.~\eqref{F-mono-chiral}. We note that following nonreciprocity of odd-elasticity, the other fifth-order stiffness tensor, $\mathbb{G}$, is fully independent of $\mathbb{F}$. 
	It is further noted that since the transformation tensor $\mathbf{R}_x(\pi)$ is purely diagonal, it only induces sign changes and never maps an off-diagonal component to its transpose. Consequently, all off-diagonal pairs in even-order stiffness tensors for monoclinic odd elastic materials remain nonreciprocal (odd).

	\subsubsection{Chiral Orthorhombic (Point Group 222)}
	\label{sec:chiralortho}
	To enforce an additional rotation symmetry, we apply a second two-fold ($180^\circ$) rotation about the $y$-axis:
	\begin{equation}
		\mathbf{R}_y(\pi) = \begin{bmatrix} -1 & 0 & 0 \\ 0 & 1 & 0 \\ 0 & 0 & -1 \end{bmatrix}.
		\label{pathA-Rot-y-180}
	\end{equation}
	Substituting Eq.~\eqref{pathA-Rot-y-180} into Eq.~\eqref{CFGA-trans} imposes the additional constraint that $N_1 + N_3$ must be even. The intersection of two orthogonal $180^\circ$ rotations about $x$ and $y$, automatically enforces rotation about $z$ as well ($\mathbf{R}_z(\pi) = \mathbf{R}_x(\pi) \mathbf{R}_y(\pi)$), enforcing that $N_1 + N_2$ must also be even. Simultaneously satisfying these conditions yield
	\begin{equation}
		\begin{aligned}
			C_{ijkl}, \, A_{ijklmn} = \;
			\begin{cases}
				\text{nonzero} & \text{if all indices appear an even number of times } (p_1=p_2=p_3=0),\\
				\text{zero} & \text{otherwise, } 
			\end{cases}
		\end{aligned}
		\label{pathA-Ortho-CA}
	\end{equation}
	and
	\begin{equation}
		\begin{aligned}
			F_{ijklm}, \, G_{ijklm} = \;
			\begin{cases}
				\text{nonzero} & \text{if all indices appear an odd number of times } (p_1=p_2=p_3=1),\\
				\text{zero} & \text{otherwise. } 
			\end{cases}
		\end{aligned}
		\label{pathA-Ortho-FG}
	\end{equation}
	
	Here, to identify the nonzero components, we partition the second- and third-order tensors into sets $\mathcal{V}_\mathbf{p}$ and $\mathcal{K}_\mathbf{p}$ based on the parity array, $\mathbf{p}$, containing parities of all indices 1, 2, and 3. The resulting sets are explicitly tabulated in Table \ref{tab:Orthorho-Sets}. Following Eq.~\eqref{pathA-Ortho-CA}, the nonzero components of even-order tensors belong to sub-blocks in which the row and column share the same parity triplet. On the other hand, Eq.~\eqref{pathA-Ortho-FG} dictates that nonzero components of odd-order tensors belong to sub-blocks in which the parities of stress and strain measures are complementary (e.g., coupling $\mathcal{V}_{000}$ with $\mathcal{K}_{111}$).
	Accordingly, using Table \ref{tab:Orthorho-Sets}, the fourth-, fifth-, and sixth-order stiffness tensors for chiral orthorhombic odd elastic materials contain $12$, $24$, and $84$ independent nonzero components, each. 
	The matrix representations of the stiffness tensors for odd elastic chiral orthorhombic materials are given in Eqs.~\eqref{Cij-ortho}-\eqref{F-222}.
	Similar to the monoclinic class, the orthogonal $\mathbf{R}_y(\pi)$ rotations do not enforce major symmetry on any surviving components.
	Consequently, the 3 off-diagonal pairs in $\mathbb{C}$ remain nonreciprocal, which increase the number of independent parameters of a chiral orthorhombic material from 9 in a conservative material to 12 in an odd elastic material. Similarly for the sixth-order tensor $\mathbb{A}$, all 33 off-diagonal components remain unequal, thereby increasing the number of independent parameters from 51 in conservative materials \cite{auffray2019complete,pirmoradi2024anisotropic} to 84 in odd elastic materials.

	From the chiral orthorhombic class, the symmetry hierarchy branches into two distinct branches, resulting in \textit{transversely hemitropic} (chiral transversely isotropic) and \textit{chiral cubic} materials.

	\subsubsection{Transversely Hemitropic (Point Group $\infty 22$)} \label{sec:TH}
	
	Transversely hemitropic (chiral transversely isotropic) symmetry group is made by adding to the symmetries of an orthorhombic material, an axis of infinite rotational symmetry. Without loss of generality, here we assume the $z$-axis to be the axis of continuous rotation symmetry, signifying that the $xy$-plane is a plane of symmetry. Enforcing the material behavior to be unchanged under this continuous rotation transformation,
	\begin{equation}
		\mathbf{R}_z(\theta) = \begin{bmatrix} \cos\theta & -\sin\theta & 0 \\ \sin\theta & \cos\theta & 0 \\ 0 & 0 & 1 \end{bmatrix},
		\label{pathA-Rot-Transverse}
	\end{equation}
	yields the following identities for the components of the fourth-order stiffness tensor:
	\begin{equation}
		\begin{aligned}
			&C_{11} = C_{22} , \quad C_{13}=C_{23}, \quad C_{31} = C_{32}, \quad C_{44} = C_{55}  \quad &&\text{(Transverse equivalence)}, \\
			&C_{66} = (C_{11} - C_{12})/2 \quad &&\text{(In-plane isotropic shear)}, \\
			&C_{12} = C_{21} \quad &&\text{(Forced in-plane $\textit{major symmetry}$)}.
		\end{aligned}
		\label{PathA-Transverse-C}
	\end{equation}
	Applying these 6 identities on the $12$ stiffness tensor components of orthorhombic materials, it is seen that the transversely hemitropic fourth-order odd elastic stiffness tensor is characterized by the following 6 independent components, $\{C_{11}, C_{33}, C_{12}, C_{13}, C_{31}, C_{44}\}$.
	While Eqs.~\eqref{PathA-Transverse-C}$_{1,2}$ contain relationships common in conservative passive transversely isotropic materials, Eq.~\eqref{PathA-Transverse-C}$_3$ reveals that imposing continuous in-plane rotational symmetry relaxes \textit{odd elasticity} in the isotropy plane. That is, even for nonconservative systems, classical major symmetry is retrieved in the in-plane direction by enforcing $ C_{12} = C_{21}$. Nevertheless, active nonconservative interactions still manifest themselves in the out-of-plane coupling where $C_{13} = C_{23} \neq C_{31} = C_{32}$. The matrix representation of the fourth-order stiffness tensor for transversely hemitropic materials is given in Eq.~\eqref{Cij-TH}.
	
	Next, substituting the rotation given in Eq.~\eqref{pathA-Rot-Transverse} into Eq.~\eqref{CFGA-trans}$_3$ results in 53 identities governing the components of the sixth-order stiffness tensor, $\mathbb{A}$. The complete system of 53 equations (comprising 30 direct transverse equalities, 13 out-of-plane shear relations, and 10 planar cross-couplings) are explicitly detailed in Eqs.~\eqref{PathA-TransverseH-A-1}-\eqref{PathA-TransverseH-A-3}. These identities reduce the number of independent components in the sixth-order stiffness tensor from 84 in the orthorhombic chiral class to 31 in the transversely hemitropic class.
	It is important to note that similar to the case of fourth-order stiffness, the continuous $\infty 22$ rotation symmetry enforces major symmetry on specific  components of the sixth-order stiffness tensor:
	\begin{equation}
		A_{10\,05} = A_{05\,10}, \quad A_{14\,13} = A_{13\,14}, \quad A_{11\,04} = A_{04\,11} \, .
		\label{PathA-Transverse-Forced-Major}
	\end{equation}
	For these specific microstructural deformation modes, even in nonconservative systems, major symmetry is recovered due to the introduction of transverse hemitropy, and the \textit{odd elasticity} effects are not maintained.
	On the other hand, 10 independent pairs can remain nonreciprocal (odd):
	\begin{equation}
		\begin{aligned}
			A_{03\,17} &\neq A_{17\,03}, \quad A_{05\,15} \neq A_{15\,05}, \quad A_{13\,15} \neq A_{15\,13}, \quad A_{01\,02} \neq A_{02\,01}, \quad A_{01\,03} \neq A_{03\,01} \\
			A_{02\,03} &\neq A_{03\,02}, \quad A_{01\,17} \neq A_{17\,01}, \quad A_{02\,17} \neq A_{17\,02}, \quad A_{05\,13} \neq A_{13\,05}, \quad A_{05\,14} \neq A_{14\,05}.
		\end{aligned}
		\label{PathA-Transverse-A-odd}
	\end{equation}
	These 10 nonreciprocal stiffness tensor components, together increase the number of independent components of $\mathbb{A}$ from 21 in conservative (even) elastic sixth-order stiffness tensors \cite{auffray2019complete, pirmoradi2024anisotropic} to 31 in odd elastic materials.
	
	Finally, substituting the rotation given in Eq.~\eqref{pathA-Rot-Transverse} into Eq.~\eqref{CFGA-trans}$_2$ yields the following identities for the components of the fifth-order stiffness tensor, $\mathbb{F}$:
	\begin{equation}
		\begin{gathered}
			\begin{alignedat}{4}
				&F_{3\,18} = 0, &\qquad &F_{6\,15} = 0, &\qquad &F_{5\,07} = -F_{4\,02}, &\qquad &F_{5\,08} = -F_{4\,01}, \\
				&F_{5\,09} = -F_{4\,03}, &\qquad &F_{1\,11} = -F_{2\,04}, &\qquad &F_{2\,11} = -F_{1\,04}, &\qquad &F_{3\,11} = -F_{3\,04}, \\
				&F_{5\,16} = -F_{4\,17}, &\qquad &F_{1\,18} = -F_{6\,13}, &\qquad &F_{2\,18} = F_{6\,13}, &\qquad &F_{6\,14} = -F_{6\,13},
			\end{alignedat} \\
			F_{5\,06} = -F_{4\,12} = (F_{4\,02} - F_{4\,01})/2, \qquad F_{6\,05} = F_{6\,10} = (F_{1\,04} - F_{2\,04})/2
		\end{gathered}
		\label{PathA-TransverseH-F}
	\end{equation}
	and similarly for $\mathbb{G}$:
	\begin{equation}
		\begin{gathered}
			\begin{alignedat}{4}
				&G_{18\,3} = 0, &\qquad &G_{15\,6} = 0, &\qquad &G_{07\,5} = -G_{02\,4}, &\qquad &G_{08\,5} = -G_{01\,4}, \\
				&G_{09\,5} = -G_{03\,4}, &\qquad &G_{11\,1} = -G_{04\,2}, &\qquad &G_{11\,2} = -G_{04\,1}, &\qquad &G_{11\,3} = -G_{04\,3}, \\
				&G_{16\,5} = -G_{17\,4}, &\qquad &G_{18\,1} = -G_{13\,6}, &\qquad &G_{18\,2} = G_{13\,6}, &\qquad &G_{14\,6} = -G_{13\,6},
			\end{alignedat} \\
			G_{06\,5} = -G_{12\,4} = (G_{02\,4} - G_{01\,4})/2, \qquad G_{05\,6} = G_{10\,6} = (G_{04\,1} - G_{04\,2})/2.
		\end{gathered}
		\label{PathA-TransverseH-G}
	\end{equation}
	Considering the additional 16 identities in either of Eqs.~\eqref{PathA-TransverseH-F} or \eqref{PathA-TransverseH-G}, the number of independent components in the fifth-order stiffness tensors reduces from 24 in orthorhombic chiral odd elasticity to 8 in transversely hemitropic odd elasticity. The matrix representation of the fifth-order stiffness tensor for transversely hemitropic materials is given in Eq.~\eqref{F-TH}.

	\subsubsection{Chiral Cubic (Point Group 23)}
	
	The chiral cubic materials branch independently from the orthorhombic symmetry and require that the material behavior in the three mutually orthogonal principal directions are identical. This is achieved by enforcing the orientation-preserving cyclic permutations of the Cartesian coordinate system ($xyz\to yzx\to zxy$), to keep the stiffness tensors unchanged.
	Accordingly, the chiral cubic class is achieved by introducing a three-fold ($120^\circ$) rotational symmetry along the $x=y=z$ axis, given by
	\begin{equation}
		\mathbf{R}_{[111]}(2\pi/3) = \begin{bmatrix} 0 & 0 & 1 \\ 1 & 0 & 0 \\ 0 & 1 & 0 \end{bmatrix},
		\label{pathA-Rot-Cubic}
	\end{equation}
	which permutes the orthogonal basis $xyz \to zxy$. Since applying this transformation twice yields the complementary $240^\circ$ rotation and maps $xyz \to yzx$, enforcing invariance under  $\mathbf{R}_{[111]}(2\pi/3)$ is sufficient to capture both permutations. 
	Substituting this transformation into Eq.~\eqref{CFGA-trans} dictates that any surviving tensor component $T_{ijk\dots}$ must equal the component obtained by the simultaneous cyclic substitution of its indices: $1 \to 3$, $2 \to 1$, and $3 \to 2$.
	Accordingly, the components of the fourth-order stiffness tensor for a cubic chiral odd elastic material must satisfy the following equalities:
	\begin{equation}
		\begin{aligned}
			&C_{11} = C_{33} = C_{22} \quad &&\text{(Main diagonal orbit)}, \\
			&C_{12} = C_{23} = C_{31} \quad &&\text{(Forward cyclic orbit)}, \\
			&C_{32} = C_{21} = C_{13} \quad &&\text{(Backward cyclic orbit)},\\
			&C_{44} = C_{66} = C_{55} \quad &&\text{(Shear orbit)}.
		\end{aligned}
		\label{PathA-Cubic-C}
	\end{equation}
	It is seen that cubic symmetry enforces 8 equality constraints, reducing the number of independent components of $\mathbb{C}$ from $12$ independent components in the orthorhombic class down to $4$. The relationships include the equality of the diagonal normal and shear components, as well as equalities in the forward and backward cycling orbits in the off-diagonal components. Notably, the terms in the forward and backward cyclic mappings, Eq.~\eqref{PathA-Cubic-C}$_{2,3}$, cross the main diagonal of the stiffness matrix. Therefore, while the cubic symmetry forces specific coupling components to be equal, odd elasticity effects are \emph{completely} preserved and the fourth-order stiffness tensor remains fully asymmetric, i.e., all nonzero off-diagonal components are independent from their transpose counterparts: $C_{12} = C_{23} = C_{31} \neq C_{21} = C_{32} = C_{13}$\footnote{We note that this survival of odd elasticity in the fourth-order cubic stiffness tensor is a unique feature of the lower-symmetry cubic point groups ($23$ and its achiral counterpart $m\bar{3}$ studied in Section \ref{subsec:achiral}). Under higher symmetric cubic classes that include 4-fold ($90^\circ$) rotations (point groups $432$ or $m\bar{3}m$), the orthogonal axes should map onto each other with a sign change as well, which would enforce $C_{12} = C_{21}$, preventing fourth-order odd elasticity \cite{rogers1963asymmetric, ostrosablin2017symmetry, yavari2025nonlinear}, as also confirmed in our expanded list of number of components for all symmetry classes in Table \ref{tab:Expanded_Summary}.}.

	Applying the same cyclic index substitution to the higher-order tensors, and utilizing the decomposition of components into different sets in Table \ref{tab:Orthorho-Sets}, yields the following equalities between the nonzero components of $\mathbb{A}$:
	\begin{equation}
		\begin{aligned}
			&A_{04\,04} = A_{11\,11} = A_{18\,18}  \quad &&\text{(Main diagonal orbit)}, \\
			&A_{04\,11} = A_{11\,18} = A_{18\,04} \quad &&\text{(Forward cyclic orbit)}, \\
			&A_{18\,11} = A_{11\,04} = A_{04\,18} \quad &&\text{(Backward cyclic orbit)}, \\
			&\mathbb{A}_{\mathcal{K}_{100}\, \mathcal{K}_{100}} = \mathbb{A}_{\mathcal{K}_{010}\, \mathcal{K}_{010}} = \mathbb{A}_{\mathcal{K}_{001}\, \mathcal{K}_{001}} \quad &&\text{(Identical } 5\times 5 \text{ sub-blocks)}.
		\end{aligned}
		\label{PathA-Cubic-A}
	\end{equation}
	Following these $56$ identities, the number of independent components of the sixth-order stiffness tensor is reduced from 84 in orthorhombic materials to $28$. We further note that, similar to the case of the fourth-order stiffness tensor, the off-diagonal terms governed by the forward and backward cyclic orbits, Eq.~\eqref{PathA-Cubic-A}$_{2,3}$, cross the matrix diagonal and therefore remain asymmetric, i.e. $A_{04\,11} = A_{11\,18} = A_{18\,04} \neq A_{11\,04} = A_{18\,11} = A_{04\,18}$. 
	Furthermore, Eq.~\eqref{PathA-Cubic-A}$_4$ equates the three $5\times 5$ sub-blocks with one another, but it does not enforce major symmetry within the block itself. Consequently, all $10$ off-diagonal pairs within this representative $5\times 5$ block remain nonreciprocal. Combined with the single nonreciprocal degree of freedom from the cyclic orbits, the cubic symmetry preserves a total of $11$ odd elastic parameters, increasing the number of independent components of cubic chiral $\mathbb{A}$ from 17 in conservative elasticity to 28 in odd elasticity.
	
	Finally, the nonzero components of the fifth-order coupling tensors, $\mathbb{F}$ and $\mathbb{G}$, of cubic chiral materials are found to satisfy the following relationships:
	\begin{equation}
		\begin{aligned}
			F_{1\,04} &= F_{3\,18} = F_{2\,11} \, , &\qquad& F_{1\,11} = F_{3\,04} = F_{2\,18} \, , \\
			F_{2\,04} &= F_{1\,18} = F_{3\,11} \, , && \mathbb{F}_{4\, \mathcal{K}_{100}} = \mathbb{F}_{5\, \mathcal{K}_{010}} = \mathbb{F}_{6\, \mathcal{K}_{001}} \, ,
		\end{aligned}
		\label{PathA-Cubic-F}
	\end{equation}
	and
	\begin{equation}
		\begin{aligned}
			G_{04\,1} &= G_{18\,3} = G_{11\,2} \, , &\qquad& G_{11\,1} = G_{04\,3} = G_{18\,2} \, , \\
			G_{04\,2} &= G_{18\,1} = G_{11\,3} \, , && \mathbb{G}_{\mathcal{K}_{100}\, 4} = \mathbb{G}_{\mathcal{K}_{010}\, 5} = \mathbb{G}_{\mathcal{K}_{001}\, 6} \, .
		\end{aligned}
		\label{PathA-Cubic-G}
	\end{equation}
	Consequently, the number of independent parameters of these stiffness tensors are reduced from $24$ in the orthorhombic class down to $8$. Matrix representations of the odd elastic stiffness tensors for chiral cubic materials are given in Eq.~\eqref{Cij-cubic}-\eqref{F-cubic}.

	\subsubsection{Hemitropic / Isotropic Chiral (Point Group $SO(3)$)}
	The chiral branch ends in the hemitropic symmetry, representing continuous rotational symmetry about all spatial axes in 3D space, $SO(3)$. This can be generated by superimposing the cubic chiral class onto the transversely hemitropic class. Effectively, by indicating that the orthogonal axes are interchangeable (i.e., cubic chiral symmetry) and simultaneously imposing continuous rotational symmetry about one axis (i.e. transversely Hemitropic symmetry), we enforce continuous rotational symmetry about all axes. 
	
	For the fourth-order stiffness tensor $\mathbb{C}$, applying this full 3D rotational symmetry yields
	\begin{equation}
		\begin{aligned}
			&C_{12} = C_{13} = C_{23} = C_{21} = C_{31} = C_{32},\\
			&C_{44} = C_{55} = C_{66} = (C_{11} - C_{12})/2.
		\end{aligned}
		\label{PathA-Hemi-C-identities}
	\end{equation}
	These relations demonstrate that the fourth-order odd elasticity tensor under hemitropic symmetry is forced to retain major symmetry and is identical to the conservative elastic isotropic stiffness tensor, formulated using Lam\'e constants as
	\begin{equation}
		C_{ijkl} = c_1 \delta_{ij} \delta_{kl} + c_2 \left( \delta_{ik} \delta_{jl} + \delta_{il} \delta_{jk} \right) \, , 
		\label{PathA-Hemi-C}
	\end{equation}
	where $\mathbf{\delta}$ denotes the Kronecker delta.
	
	For the sixth-order tensor $\mathbb{A}$, superposing the conditions of cubic chiral symmetry, Eq.~\eqref{PathA-Cubic-A}, onto those of transversely hemitropic symmetry, Eqs.~\eqref{PathA-TransverseH-A-1}-\eqref{PathA-TransverseH-A-3}, reduces the number of independent components to 6. Comparing this to the classical conservative sixth-order elasticity tensor \cite{mindlin1963microstructure,dell2009generalized,pirmoradi2024anisotropic}, which contains 5 independent parameters, reveals that odd elasticity splits the first higher-order parameter $a_1$ into two distinct nonreciprocal moduli, $a_{1a}$ and $a_{1b}$. The resulting hemitropic odd elasticity tensor can be written as a linear combination of Kronecker deltas:
	\begin{equation}
		\begin{array}{lcl}
			A_{ijklmn}& = & a_{1a} (\delta_{ij}\delta_{kl}\delta_{mn}+\delta_{ik}\delta_{jl}\delta_{mn}) \ + \ a_{1b} (\delta_{im}\delta_{jk}\delta_{ln}+\delta_{in}\delta_{jk}\delta_{lm}) \\ 
			& + & a_2(\delta_{ij}\delta_{km}\delta_{ln}+\delta_{ij}\delta_{kn}\delta_{lm}+\delta_{ik}\delta_{jm}\delta_{ln}+\delta_{ik}\delta_{jn}\delta_{lm})\\
			& + & a_3(\delta_{il}\delta_{jk}\delta_{mn})\\
			& + & a_4(\delta_{il}\delta_{jm}\delta_{kn}+\delta_{il}\delta_{jn}\delta_{km})\\
			& + & a_5(\delta_{im}\delta_{jl}\delta_{kn}+\delta_{im}\delta_{jn}\delta_{kl}+\delta_{in}\delta_{jl}\delta_{km}+\delta_{in}\delta_{jm}\delta_{kl}) \, . 
		\end{array}
		\label{PathA-Hemi-A}
	\end{equation}
	We note that the basis tensors associated with $a_{1a}$ and $a_{1b}$ are major transposes of one another. Accordingly, by splitting the two terms, major symmetry is revoked to accommodate odd elasticity. It is crucial to note that this split also retains minor symmetries of stiffness tensor, which is required to keep the balance of angular momentum and objectivity \cite{yavari2025nonlinear, wolfgram2025odd}.
	
	Finally, superimposing the conditions of cubic chiral symmetry of the fifth-order elasticity tensors, Eq.~\eqref{PathA-Cubic-F}, onto the conditions of transversely hemitropic symmetry, Eq.~\eqref{PathA-TransverseH-F} yields the identities required for the chiral hemitropic class. For this case, the fifth-order elasticity tensor in the conservative elastic limit can be expressed by products of the Kronecker delta and the Levi-Civita permutation tensor. As shown in \cite{papanicolopulos2011chirality, auffray2019complete}, imposing minor symmetries, reduces the combinations into a single parameter for each of the fifth-order tensors. Revoking Maxwell-Betti reciprocity for nonconservative materials simply requires independence of the two coupling tensors. Accordingly, the tensor $\mathbb{F}$ reads:
	\begin{equation}
		F_{klmij} = f_1 \left( \epsilon_{kli}\delta_{mj} + \epsilon_{kmi}\delta_{lj} + \epsilon_{klj}\delta_{mi} + \epsilon_{kmj}\delta_{li} \right) \, ,
		\label{PathA-Hemi-F}
	\end{equation}
	and $\mathbb{G}$ reads:
	\begin{equation}
		G_{ijklm} = g_1 \left( \epsilon_{ijl}\delta_{km} + \epsilon_{ikl}\delta_{jm} + \epsilon_{ijm}\delta_{kl} + \epsilon_{ikm}\delta_{jl} \right) \, .
		\label{PathA-Hemi-G}
	\end{equation}
	We note that the basis tensors in Eqs.~\eqref{PathA-Hemi-F} and \eqref{PathA-Hemi-G} are major transposes of one another; therefore, the classical conservative limit (established by imposing this reciprocity via $\mathbb{G}=\mathbb{F}^\mathsf{T}$) requires $g_1 = f_1$. Matrix representation of fourth- and fifth-order odd elasticity tensors of hemitropic materials are given in Eq.~\eqref{Cij-iso},\eqref{F-SO3}.
	
	\subsection{The Achiral Branch (Improper Rotations)}
	\label{subsec:achiral}
	In this branch, elastic symmetries are initiated by incorporating planes of elastic symmetry \cite{Malvern-Continuum} and finally arrive at fully isotropic materials, which have the entire orthogonal transformations as their symmetry group \cite{love1892treatise}.
	It is noted that reflection transformation about any plane can be obtained by applying a central inversion ($-\mathbf{I}$) to the two-fold rotation transformation about the normal axis of the same plane. Since for even-order tensors, central inversion amounts to multiplying the tensor with $-1$ an even number of times, and therefore, leaves the tensor unchanged, the symmetry classes obtained in the chiral group for the fourth- and sixth-order tensors, $\mathbb{C}$ and $\mathbb{A}$, can be used identically for the achiral group in this section. Accordingly, in this section we only study the effects of applied elastic symmetries on fifth-order stiffness tensors.
	
	\subsubsection{Achiral Monoclinic (Point Group $m$)}
	
	In these materials, the symmetry group consists of a single plane of reflection symmetry. Without loss of generality, and to maintain consistency with the chiral monoclinic analysis in Section \ref{sec:chiralmono}, we assume the $yz$-plane to be the plane of elastic symmetry. Therefore the elasticity tensors need to remain unchanged under the following transformation of the coordinate system
	\begin{equation}
		\mathbf{M}_{yz} = -\mathbf{R}_x(\pi) = \begin{bmatrix} -1 & 0 & 0 \\ 0 & 1 & 0 \\ 0 & 0 & 1 \end{bmatrix}.
		\label{pathB-Reflect-yz}
	\end{equation}
	As established earlier, the conditions for the fourth- and sixth-order elasticity tensors of achiral monoclinic materials are identical to those demonstrated in Section \ref{sec:chiralmono}. 
	On the other hand, for the fifth-order elasticity tensors, the conditions for the nonzero components are the inverse of those derived for chiral monoclinic materials, and therefore read
	\begin{equation}
		\begin{aligned}
			F_{ijklm}, \, G_{ijklm} = \;
			\begin{cases}
				\text{nonzero} & \text{if index 1 appears an even number of times ($p_1=0$),} \\
				\text{zero} & \text{if index 1 appears an odd number of times ($p_1=1$). } 
			\end{cases}
		\end{aligned}
		\label{pathB-Mono-FG}
	\end{equation}
	Matrix representation of this stiffness tensor for achiral monoclinic nonconservative materials is given in Eq.~\eqref{F-mono-achiral} and shows that $\mathbb{F}$ and $\mathbb{G}$ contain $56$ independent nonzero components, each, which are the complementary set of the $52$ nonzero components of the chiral monoclinic case, Eq.~\eqref{F-mono-chiral}.

	\subsubsection{Achiral Orthorhombic (Point Group $mm2$)}
	
	By introducing a second orthogonal plane of reflection, we arrive at the achiral orthorhombic class. Without loss of generality, we assume the $xz$-plane to be the second plane of elastic symmetry, with its corresponding transformation tensor given by:
	\begin{equation}
		\mathbf{M}_{xz} = -\mathbf{R}_y(\pi)= \begin{bmatrix} 1 & 0 & 0 \\ 0 & -1 & 0 \\ 0 & 0 & 1 \end{bmatrix}.
		\label{pathB-Reflect-xz}
	\end{equation}
	The conditions for the even-order tensors ($\mathbb{C}$ and $\mathbb{A}$) remain identical to those of the chiral orthorhombic class in Eq.~\eqref{pathA-Ortho-CA}.
	However, for the fifth-rank tensors, substituting Eq.~\eqref{pathB-Reflect-xz} into Eq.~\eqref{CFGA-trans}$_{2,3}$ and superimposing the result onto the achiral monoclinic condition, Eq.~\eqref{pathB-Mono-FG}, shows that the nonzero components of $\mathbb{F}$ and $\mathbb{G}$ have an even number of indices $1$ and $2$, and an odd number of index $3$:
	\begin{equation}
		\begin{aligned}
			F_{ijklm}, \, G_{ijklm} = \;
			\begin{cases}
				\text{nonzero} & \text{if the total parity array is } \mathbf{p} = (0,0,1), \\
				\text{zero} & \text{otherwise. } 
			\end{cases}
		\end{aligned}
		\label{pathB-Ortho-FG}
	\end{equation}
	The resulting matrix representation of the fifth-order stiffness tensor for achiral orthorhombic materials is shown in Eq.~\eqref{F-mm2} and show that the coupling tensors $\mathbb{F}$ and $\mathbb{G}$ contain $28$ independent components, each.

	\subsubsection{Orthotropic (Point Group $mmm$) and Higher Elastic Symmetry Levels}
	\label{sec:OrthoAndHigher}
	
	Orthotropic tensors are those with three mutually orthogonal planes of reflection symmetry. This is achieved by adding to Eqs. \eqref{pathB-Reflect-yz} and \eqref{pathB-Reflect-xz}, a third reflection symmetry, with its corresponding transformation tensor
	\begin{equation}
		\mathbf{M}_{xy} = \mathbf{R}_z(\pi) = \begin{bmatrix} 1 & 0 & 0 \\ 0 & 1 & 0 \\ 0 & 0 & -1 \end{bmatrix}.
		\label{pathB-Reflect-xy}
	\end{equation}
	We note here that the sequential application of reflections across three mutually-orthogonal planes is identical to a global central inversion, as $\mathbf{M}_{xy} \mathbf{M}_{xz} \mathbf{M}_{yz} = -\mathbf{I}$. Consequently, the $mmm$ symmetry group inherently contains a center of inversion, establishing centrosymmetry \cite{lakes1982noncentrosymmetry,lakes2001elastic,pirmoradi2024anisotropic}. 
	As established earlier, the fourth- and sixth-order odd elasticity tensors for orthotropic symmetry ($mmm$) are identical to the chiral orthorhombic class, Eqs.~\eqref{Cij-ortho} and \eqref{A-ortho}.
	On the other hand, the application of central inversion on fifth-order tensors amounts to multiplying the tensor with $-1$ an odd number of times, thereby negating the tensor. Accordingly, the fifth-order stiffness tensors for odd elastic centrosymmetric materials (including orthotropic materials), fully vanish, $\mathbb{F} = \mathbb{G} = \mathbf{0}$.
	
	These conditions propagate to higher levels of elastic symmetry in the achiral group as well. Consequently, the even-order stiffness tensors for transversely isotropic materials (Point Group $\infty/mmm$), achiral cubic materials (Point Group $m\bar{3}$), and isotropic materials (Point Group $O(3)$) are identical to those belonging to transversely hemitropic materials (Point Group $\infty 22$), chiral cubic materials (Point Group $23$), and hemitropic materials (Point Group $SO(3)$), respectively.
	Similarly, the fifth-order stiffness tensors for all aforementioned achiral symmetry groups vanish.

	Table~\ref{tab:Master_Summary} demonstrates a summary of the number of independent components of stiffness tensors for the different levels of anisotropy studied here. To explicitly highlight the effects of relaxing the Maxwell-Betti reciprocity, thereby introducing odd elasticity, the values for conservative (major-symmetric) tensors are also presented inside parentheses. As suggested by Yavari and Goriely for the first-gradient framework \cite{yavari2025nonlinear}, due to the nonconservative nature of odd elasticity, experimental application of cyclic deformation paths can isolate the effects of, and reveal the values of antisymmetric stiffness components. Similar loading scenarios need to be formulated and can be simulated on strain-gradient odd elastic systems, following numerical micromechanical simulations of higher-gradient conservative systems by the author, \cite{Payam-Identification, misra2017elastic,poorsolhjouy2018connecting}.

	\begin{table}[ht]
		\centering
		\renewcommand{\arraystretch}{1.4}
		\begin{tabular}{l c | c c c}
			\hline
			\multicolumn{2}{c|}{\textbf{Symmetry Class}} & \multicolumn{3}{c}{\textbf{Stiffness Tensors}} \\
			\hline
			\hline
			\textit{Chiral materials} & \textbf{Point Group} & $\mathbb{C}$ (4th order) & $\mathbb{F}, \mathbb{G}$ (5th order) & $\mathbb{A}$ (6th order) \\
			\hline
			Triclinic & 1 & 36 (21) & $2\times 108$ (108) & 324 (171) \\
			Monoclinic & 2 & 20 (13) & $2\times 52$ (52) & 164 (91) \\
			Orthorhombic & 222 & 12 (9) & $2\times 24$ (24) & 84 (51) \\
			Transversely Hemitropic & $\infty 22$ & 6 (5) & $2\times 8$ (8) & 31 (21) \\
			Cubic & 23 & 4 (3) & $2\times 8$ (8) & 28 (17) \\
			Hemitropic (Isotropic) & $SO(3)$ & 2 (2) & $2\times 1$ (1) & 6 (5) \\
			\hline
			\textit{Achiral materials} & \textbf{Point Group} & $\mathbb{C}$ (4th order) & $\mathbb{F}, \mathbb{G}$ (5th order) & $\mathbb{A}$ (6th order) \\
			\hline
			Monoclinic & $m$ & 20 (13) & $2\times 56$ (56) & 164 (91) \\
			Orthorhombic & $mm2$ & 12 (9) & $2\times 28$ (28) & 84 (51) \\
			Orthotropic (Centrosymmetric) & $mmm$ & 12 (9) & 0 (0) & 84 (51) \\
			Transversely Isotropic & $\infty/mmm$ & 6 (5) & 0 (0) & 31 (21) \\
			Cubic & $m\bar{3}$ & 4 (3) & 0 (0) & 28 (17) \\
			Isotropic & $O(3)$ & 2 (2) & 0 (0) & 6 (5) \\
			\hline
		\end{tabular}
		\caption{Number of independent components for the strain-gradient odd elasticity tensors. Values are presented as $N (N_{\text{classical}})$, where $N$ is the number of independent parameters in the nonconservative odd elastic framework, and $N_{\text{classical}}$ is the number of parameters in classical conservative elasticity. For the fifth-order tensors, the notation $2\times N (N)$ highlights that $\mathbb{F}$ and $\mathbb{G}$ are decoupled in odd elasticity (yielding $2N$ total parameters), whereas classical elasticity enforces $\mathbb{G} = \mathbb{F}^{\mathsf{T}}$ (yielding $N$ total parameters).}
		\label{tab:Master_Summary}
	\end{table}

	\section{Governing Equations and Non-Hermitian Wave Propagation}
	\label{sec:waves}
	
	In this section, we derive the equations of motion and boundary conditions of strain-gradient odd elasticity and demonstrate the effect of relaxing the Maxwell-Betti reciprocity on the elastodynamic response, leading to non-Hermitian acoustic tensor.

	\subsection{Equations of Motion and Boundary Conditions}
	We establish the dynamic equilibrium equations for the higher-order continuum using the dynamic principle of virtual work, which requires the total virtual work done by the external forces on the system to be equal to the sum of the internal virtual work and the virtual work of the inertial forces for any arbitrary admissible virtual displacement:
	\begin{equation}
		\delta W^{\text{ext}} = \delta W^{\text{int}} +\delta W^{\text{iner}} \, .
		\label{Eq-Virtual-Work}
	\end{equation}
	The internal virtual work is defined as the integral of the inner products of the virtual kinematic measures and their work-conjugate stress measures:
	\begin{equation}
		\delta W^{\text{int}} = \int_V \left( \tau_{ij} \delta u_{i,j} + \mu_{ijk} \delta u_{i,jk} \right) dV \, ,
	\end{equation}
	where the symmetry of the Cauchy stress ($\tau_{ij} = \tau_{ji}$) allows the use of the displacement gradient instead of the strain tensor. It is crucial to emphasize that in the context of odd elasticity and in the absence of a strain energy potential, $\delta W^{\text{int}}$ represents an inexact differential of virtual work rather than the variation of the strain energy potential (a state function). The ability of the active system to inject or extract net mechanical work over a closed deformation cycle simply means that this internal work is path-dependent.

	Applying integration by parts once to the first term and twice to the second term of the integral above yields:
	\begin{equation}
		\delta W^{\text{int}} =     \int_{\partial V} \left[\left(\tau_{ij}-\mu_{ijk,k}\right) n_j \delta u_i + \mu_{ijk} n_k \delta u_{i,j}\right]\, dS 
		+ \int_V \left(\mu_{ijk,jk} - \tau_{ij,j}\right)\delta u_i \, dV \, ,
		\label{Wint-1}
	\end{equation}
	with $\mathbf{n}$ the outward unit normal to the boundary.
	Following Mindlin's framework of strain-gradient elasticity \cite{mindlin1965second}, it is noted that the gradient of the virtual displacement, $\delta u_{i,j}$, on the boundary is not independent of the displacement field within the volume. Accordingly, the gradient of displacement is orthogonally decomposed into its surface gradient (fully determined by the displacement field within the volume) and its normal derivative (independent of the displacement within the volume) components, as
	\begin{equation}
		\delta u_{i,j} = D_j \delta u_i + n_j D_\text{n} \delta u_i \,,
		\label{DnDj}
	\end{equation}
	where $D_j = (\delta_{jl} - n_j n_l)\partial_l$ and $D_\text{n} = n_l \partial_l$ are the surface gradient  and the normal derivative operators, respectively \cite{toupin1962elastic,mindlin1965second,eremeyev2024effective}. 
	
	Since the internal virtual work is based on two independent kinematic variations (the displacement field across the body volume $\delta \mathbf{u}$ and its normal derivative on the boundary $D_\text{n} \delta \mathbf{u}$), the external virtual work must also postulate conjugate force measures for both. 
	Therefore, $\delta W^{\text{ext}}$ includes contributions from the body force $\mathbf{f}$ in the bulk, the standard boundary traction $\mathbf{t}$ (conjugate to $\delta \mathbf{u}$), and the higher-order double-traction $\mathbf{r}$ (conjugate to $D_\text{n} \delta \mathbf{u}$):
	\begin{equation}
		\delta W^{\text{ext}} = \int_V f_i \delta u_i \, dV + \int_{\partial V} \left( t_i \delta u_i + r_i D_\text{n} \delta u_i \right) dS \, .
		\label{Wext}
	\end{equation}
	In classical conservative strain-gradient elasticity, this incorporation of bulk and boundary forces ensures retaining Maxwell-Betti reciprocity \cite{xu2024does}.
	Finally, the virtual work of the inertial forces is given by:
	\begin{equation}
		\delta W^{\text{iner}} = \int_V \rho \ddot{u}_i \delta u_i \, dV \, ,
		\label{Winner}
	\end{equation}
	where $\rho$ is the mass density and superimposed dots denote time derivatives. 
	Substituting Eq.~\eqref{Wint-1} (updated via the decomposition in Eq.~\eqref{DnDj}), along with Eqs.~\eqref{Wext}, and \eqref{Winner} back into the virtual work balance, Eq.~\eqref{Eq-Virtual-Work}, and enforcing that the relationship is satisfied for arbitrary kinematic variations, yields the higher-order equations of motion in the bulk:
	\begin{equation}
		\tau_{ij,j} - \mu_{ijk,jk} + f_i = \rho \ddot{u}_i \, ,
		\label{Equil}
	\end{equation}
	along with the boundary conditions:
	\begin{equation}
		\begin{aligned}
			t_i &= \tau_{ij} n_j - \mu_{ijk,k} n_j - D_j(\mu_{ijk} n_k) + \mu_{ijk} n_j n_k (D_l n_l) \, , \\
			r_i &= \mu_{ijk} n_j n_k \, .
		\end{aligned}
		\label{Eq-Boundary-Conditions}
	\end{equation}

	It is crucial to note that because the dynamic principle of virtual work is a statement of mechanical momentum balance, it is completely independent of the constitutive laws that reflect the passive/active nature of the system. Consequently, the derived equilibrium equations and boundary conditions for odd elastic second-gradient continua are identical to those of classical conservative strain-gradient elasticity \cite{mindlin1965second}.

	\subsection{Equations of Motion and Elastic Wave Propagation}
	The nonconservative nature of the odd elastic continuum manifests itself when the nonreciprocal constitutive relations, Eq.~\eqref{Const-Macro}, are substituted into the equilibrium equations, Eq.~\eqref{Equil}, and result in the generalized displacement equations of motion. In the absence of body forces, the displacement equations of motion for a generalized odd elastic strain-gradient continuum read:
	\begin{equation}
		C_{ijkl} u_{k,lj} + \left( F_{ijklm} - G_{ijlkm} \right) u_{k,lmj} - A_{ijlkmn} u_{k,mnjl} = \rho \ddot{u}_i \, .
		\label{Eq-Navier-Cauchy}
	\end{equation}
	We note that in the limit of classical conservative elasticity, where the existence of a strain energy potential enforces Maxwell-Betti reciprocity and ensures the higher-order coupling tensors to be transposes of each other ($F_{ijklm} = G_{klmij}$), Eq.~\eqref{Eq-Navier-Cauchy} reduces to the classical displacement equations of motion for conservative Form II strain-gradient materials, as originally derived by Mindlin \cite{mindlin1965second, mindlin1968first}.

	To analyze wave dispersion inside odd elastic materials, we assume a time-harmonic plane wave solution of the form:
	\begin{equation}
		\mathbf{u}(\mathbf{x}, t) = \mathbf{U} e^{\mathrm{i} (\mathbf{k} \cdot \mathbf{x} - \omega t)} \, ,
		\label{Eq-Plane-Wave}
	\end{equation}
	where $\mathbf{U}$ is the amplitude vector, $\mathbf{k}$ is the wave vector, and $\omega$ is the angular frequency. 
	Substituting this harmonic wave into the equations of motion, Eq.~\eqref{Eq-Navier-Cauchy}, yields the following eigenvalue problem governing wave propagation:
	\begin{equation}
		\left(\Gamma_{ik} - \rho \omega^2 \delta_{ik}\right) U_k =  0 \, ,
		\label{eigval}
	\end{equation}
	where $\Gamma_{ik}$ denotes the components of the generalized acoustic tensor, which, similar to the case of classical conservative elasticity, can be decomposed into its real and imaginary parts as
	\begin{equation}
		\Gamma_{ik} = \underbrace{C_{ijkl} k_j k_l + A_{ijlkmn} k_j k_l k_m k_n}_{\Gamma_{ik}^\text{R}} + \mathrm{i} \underbrace{\left( F_{ijklm} - G_{ijlkm} \right) k_j k_l k_m}_{\Gamma_{ik}^\text{I}} \, .
		\label{Eq-Acoustic-Tensor}
	\end{equation}
	For nontrivial wave solutions to exist ($\mathbf{U} \neq \mathbf{0}$), the determinant of the characteristic matrix must vanish, yielding the characteristic equation $\det(\Gamma_{ik} - \rho \omega^2 \delta_{ik}) = 0$.
	
	It is noted that, similar to conservative elasticity, the real and imaginary parts of the acoustic tensor are constructed from even and odd powers of wave vector, and are therefore, even and odd functions, respectively:
	\begin{equation}
		\mathbf{\Gamma}^\text{R}(-\mathbf{k}) = \mathbf{\Gamma}^\text{R}(\mathbf{k}) \, , \qquad \mathbf{\Gamma}^\text{I}(-\mathbf{k}) = -\mathbf{\Gamma}^\text{I}(\mathbf{k}) \, .
		\label{QrQi-EvenOdd}
	\end{equation}
	It is, therefore, seen that even in strain-gradient odd elasticity, the acoustic tensor for each propagation direction is the complex conjugate of its reverse direction: $\mathbf{\Gamma}(-\mathbf{k}) = \bar{\mathbf{\Gamma}}(\mathbf{k})$. Consequently, the eigenvalues of a wave propagating in a given direction are also the complex conjugates of those propagating in the reverse direction: $\omega^2(-\mathbf{k}) = \bar{\omega}^2(\mathbf{k})$. In other words, even in strain-gradient odd elasticity, if a wave propagating in $\mathbf{k}$ has an eigenvector $\mathbf{v}$ and a frequency eigenvalue $\omega^2$, the same wave propagating in $-\mathbf{k}$ will have the conjugated eigenvector $\bar{\mathbf{v}}$ and the conjugated frequency eigenvalue $\bar{\omega}^2$.

	The departure of odd elasticity from conservative mechanics in elastodynamics manifests in the Hermiticity of the acoustic tensor. For a wave system to be conservative (exhibiting purely real frequencies, $\omega^2 \in \mathbb{R}$), the acoustic tensor must be Hermitian ($\mathbf{\Gamma} = \bar{\mathbf{\Gamma}}^\mathsf{T}$). 
	In Appendix \ref{app:hermiticity}, two independently sufficient conditions for the Hermiticity of $\mathbf{\Gamma}$ are demonstrated: (i) the enforcement of classical major symmetries of passive conservative materials ($C_{ijkl}=C_{klij}$, $A_{ijklmn} = A_{lmnijk}$, and $F_{ijklm} = G_{klmij}$), or (ii) the enforcement of isotropic or hemitropic (isotropic chiral) elastic symmetries ($O(3)$ or $SO(3)$)--this is despite the facts that for nonconservative odd elasticity (a) even in the isotropic limit, the sixth-order stiffness tensor retains its major asymmetry through the split of the odd elastic stiffness coefficients $a_{1a} \neq a_{1b}$, and (b) in the hemitropic case, the fifth-order stiffness coefficients $f_1$ and $g_1$ are nonzero and independent. 
	However, for odd elastic materials with higher levels of anisotropy, both of these sufficient conditions are violated and the asymmetric constitutive moduli may survive the wave vector contractions into acoustic tensor. In this condition, the acoustic tensor becomes non-Hermitian and allows complex frequency eigenvalues ($\omega^2 \notin  \mathbb{R}$).

	\section{Results: Elastodynamics in Odd Elastic Media}
	\label{sec:results}
	
	To explicitly demonstrate the physical impact of odd elasticity on wave dispersion, we solve the equations of motion for elastic plane waves in nonconservative active materials with odd elasticity and different levels of anisotropy. We first analyze a fully isotropic, centrosymmetric medium to isolate the effect of survival of sixth-order odd elasticity, and subsequently extend the analysis to a hemitropic medium to demonstrate chiral wave splitting. Finally, to demonstrate the ability of odd elasticity to make the acoustic tensor non-Hermitian, and its effect on wave propagation, we analyze elastic wave propagation within an odd elastic transversely hemitropic (chiral) material.
	
	As established by Scheibner et al. \cite{scheibner2020odd} for first-gradient media, odd elasticity allows a material to inject or extract net mechanical work during a closed deformation cycle. In the context of the strain-gradient elasticity studied here, a net nonzero work over a closed quasi-static elastic cycle is formulated as $\oint (\tau_{ij} \mathrm{d}\epsilon_{ij} + \mu_{ijk} \mathrm{d}\kappa_{ijk}) \neq 0$. Since a time-harmonic plane wave subjects the local material to a closed kinematic cycle, the nonconservative nature of the odd elastic moduli can manifest dynamically as active energy exchange within elastic wave propagation. Considering the displacement of the wave given in Eq.~\eqref{Eq-Plane-Wave}, the presence of complex frequency eigenvalues dictates that the wave oscillates while simultaneously demonstrating active amplification or attenuation (based on the sign of the imaginary component), signifying continuous active energy exchange over closed kinematic cycles.

	\subsection{Wave Propagation in an Odd Elastic Isotropic Material}
	\label{sec:res_iso}
	
	Since in isotropic materials, the elastic behavior is independent of wave propagation directions, without loss of generality we focus on a plane wave propagating along the $x$-axis. In this case, the wave vector is $\mathbf{k} = (k, 0, 0)^\mathsf{T}$, and the generalized acoustic tensor, Eq.~\eqref{Eq-Acoustic-Tensor}, simplifies to
	\begin{equation}
		\Gamma_{ik} = C_{i1k1} k^2 + A_{i11k11} k^4 \, .
		\label{Q-x1-iso}
	\end{equation}
	We note that due to centrosymmetry, the fifth-order stiffness tensors have vanished, $\mathbb{F} = \mathbb{G} = \mathbf{0}$.
	Utilizing the constitutive tensors given in Eqs.~\eqref{PathA-Hemi-C} and \eqref{PathA-Hemi-A}, and substituting the acoustic tensor above into the eigenvalue equation of motion, Eq.~\eqref{eigval}, the condition for wave propagation in the $x$-direction in an odd elastic isotropic material is derived as:
	
	\begin{equation}
		\begin{vmatrix}
			C_\text{L}k^2 + A_\text{L} k^4 - \rho \omega^2& 0 & 0 \\
			0 & C_\text{S} k^2 + A_\text{S} k^4 - \rho \omega^2 & 0 \\
			0 & 0 & C_\text{S} k^2 + A_\text{S} k^4 - \rho \omega^2
		\end{vmatrix} 
		= 0.
		\label{determ-x1-iso}
	\end{equation}
	where $C_\text{S}$ and $C_\text{L}$ are the standard longitudinal and shear stiffness coefficients,
	\begin{equation}
		C_\text{L} \equiv C_{1111} = c_1 + 2c_2 \, , \quad C_\text{S} \equiv C_{2121} = C_{3131} = c_2 \, ,
		\label{CSCL}
	\end{equation}
	and $A_\text{L}$ and shear $A_\text{S}$ are the longitudinal and shear gradient stiffness coefficients,
	\begin{equation}
		\begin{aligned}
			A_\text{L} \equiv A_{111111} &= 2\left(a_{1a} + a_{1b}\right) + 4a_2 + a_3 + 2a_4 + 4a_5  \, , \\
			A_\text{S} \equiv A_{211211} &= A_{311311} = a_3 + 2a_4 \, .
		\end{aligned}
		\label{ASAL}
	\end{equation}
	The fact that the acoustic tensor is diagonal with real values confirms the analytical conclusion in Appendix \ref{app:hermiticity} that the acoustic tensor for isotropic strain-gradient odd elasticity remains unconditionally Hermitian. Consequently, the wave propagation remains strictly conservative (yielding purely real frequency eigenvalues, $\omega^2 \in \mathbb{R}$), and the nonconservative nature of the active interactions is elastodynamically hidden.
	
	Solving the eigenvalue problem above yields two distinct roots corresponding to the longitudinal and transverse components of the wave (we note that in line with isotropy, the two transverse waves are identical as $\Gamma_{22} = \Gamma_{33}$):
	\begin{equation}
		\begin{aligned}
			&\omega_\text{L} =  \sqrt{\frac{C_\text{L}k^2 + A_\text{L} k^4}{\rho}} \,, \qquad &\text{Longitudinal component},\\
			&\omega_\text{T} =  \sqrt{\frac{C_\text{S} k^2  + A_\text{S} k^4 }{\rho}} \,, \qquad &\text{Transverse components}.
		\end{aligned}
		\label{disp-iso}
	\end{equation}
	For our numerical example, we adopt the properties of an isotropic representative volume element (RVE) as the baseline conservative properties. Specifically, the parameters are extracted from a previous study that uses discrete and granular micromechanics simulations of a random assembly of 576 grains with diameters ranging from 19 $\mu$m to 28 $\mu$m \cite{misra2017elastic}. The homogenized macroscopic stiffness components of this assembly and their mapping to the classical conservative isotropic moduli ($c_1$ and $c_2$ for the fourth-order tensor $\mathbb{C}$, and $a_1, \dots, a_5$ for the sixth-order tensor $\mathbb{A}$) are summarized in Table~\ref{tab:baseline_params}.

	\begin{table}[h]
		\centering
		\renewcommand{\arraystretch}{1.4}
		\resizebox{\textwidth}{!}{%
			\begin{tabular}{cc | cc | ccccc | ccccc}
				\hline
				\multicolumn{4}{c|}{\textbf{Fourth-order ($\mathbb{C}$) [GPa]}} & \multicolumn{10}{c}{\textbf{Sixth-order ($\mathbb{A}$) [N]}} \\
				\hline\hline
				\multicolumn{2}{c|}{Tensor Components} & \multicolumn{2}{c|}{Isotropic Moduli} & \multicolumn{5}{c|}{Tensor Components} & \multicolumn{5}{c}{Isotropic Moduli} \\
				\hline
				$C_{1111}$ & $C_{1122}$ & $c_1$ & $c_2$ & $A_{111111}$ & $A_{211211}$ & $A_{112112}$ & $A_{122122}$ & $A_{111122}$ & $a_1$ & $a_2$ & $a_3$ & $a_4$ & $a_5$ \\
				$42.6$ & $1.6$ & $1.6$ & $20.5$ & $2.53$ & $2.54$ & $1.28$ & $2.54$ & $-0.03$ & $0.6275$ & $-0.6425$ & $1.2550$ & $0.6425$ & $0.0125$ \\
				\hline
			\end{tabular}%
		}
		\caption{Baseline conservative material parameters extracted from discrete and granular micromechanics simulations of an isotropic granular assembly \cite{misra2017elastic}.}
		\label{tab:baseline_params}
	\end{table}

	To study the effects of odd elastic parameters, we phenomenologically superimpose nonreciprocal perturbations onto this conservative baseline. As established in the literature of odd elasticity, e.g., \cite{brandenbourger2019non, scheibner2020odd, chen2021realization}, the introduction of nonreciprocal active feedback loops in structured media result in macroscopic asymmetric stiffness moduli. However, as demonstrated in Section \ref{sec:OrthoAndHigher} and Eq.~\eqref{PathA-Hemi-A}, in the limit of fully isotropic symmetry, this will only manifest in the sixth-order stiffness tensor by splitting the classical parameter $a_1$ into nonreciprocal components $a_{1a} \neq a_{1b}$. In this numerical demonstration, we have kept $a_{1a} = a_1$, but perturbed its nonreciprocal counterpart to $a_{1b} = 1.5a_1$, with the magnitude of the perturbation reflecting the active nonreciprocal energy injection relative to the passive structural stiffness.

	Using these mechanical properties, the intrinsic material scaling parameters, i.e. the microstructural length scale of the continuum, $\ell$, and the reference shear frequency, $\omega_0$, are defined as
	\begin{equation}
		\ell \equiv \sqrt{A_\text{S}/C_\text{S}}\  ,\quad 
		\omega_0 \equiv \ell^{-1}\sqrt{C_\text{S}/\rho},
	\end{equation}
	where $\rho$ is the mass density of the granular system, for which a typical value of $2000$ kg/m$^3$ is assumed. The length scale, $\ell$, denotes the characteristic size of the underlying heterogeneities and dictates the spatial scale at which nonlocal gradient effects become significant \cite{mindlin1965second, askes2011gradient}. Further, the reference shear frequency, $\omega_0$, represents the characteristic frequency at which the wavelength of a propagating wave approaches the microstructural length scale $\ell$, \cite{papargyri2009wave}. 
	
\begin{figure}[h]
		\centering
		\includegraphics[width=0.7\textwidth]{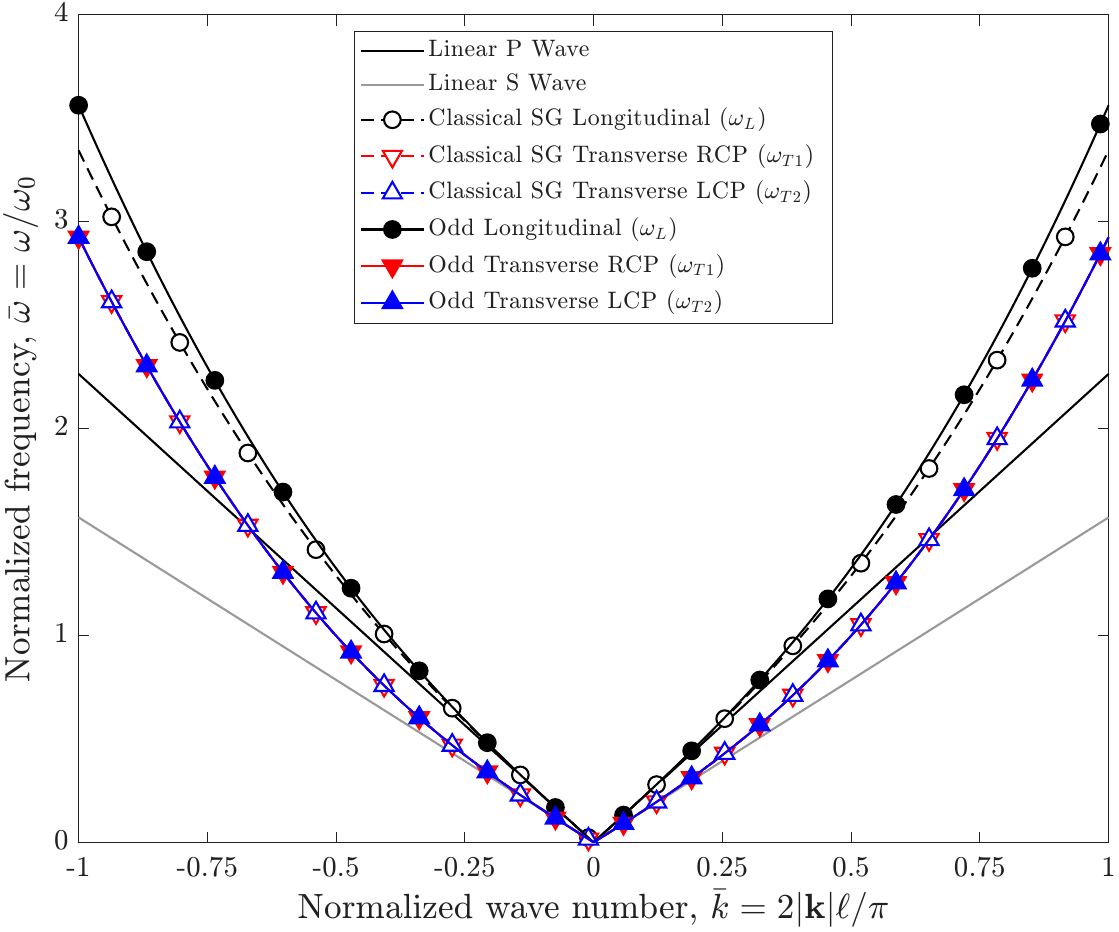}
		\caption{Normalized dispersion curves for a plane wave propagating in an isotropic odd elastic continuum (solid lines with solid markers). The linear acoustic asymptotes (lines) and classical conservative strain-gradient limits (dashed lines with hollow markers) are provided for reference.}
		\label{fig:dispersion_iso}
	\end{figure}

	In Figure \ref{fig:dispersion_iso} we have presented the dispersion curves for the odd elastic isotropic material parameters defined, with normalized wave number and frequency as $\bar{k} = 2k\ell/\pi$ and $\bar{\omega} = \omega/\omega_0$. To also study the effects of inverting the propagation direction on the acoustic tensor, Eq.~\eqref{QrQi-EvenOdd}, all dispersion curves are plotted for both forward and backward propagation, and in the wave length domain bounded by $\bar{k}\in[-1,+1]$ \cite{brillouin-Book,Sui2001a,Sui2001b}.
	To explicitly highlight the impact of odd elasticity, the plot includes two classical limits for comparison: (i) the nondispersive linear acoustic waves (P and S waves) derived from first-gradient analysis, shown as straight lines, and (ii) the classical conservative strain-gradient waves (where reciprocity enforces $a_{1b} = a_{1a}$), shown as dashed curves with hollow markers.

	We note from Figure \ref{fig:dispersion_iso} that the introduction of odd elasticity modifies the longitudinal component of the elastic wave compared to the classical conservative second-gradient baseline. As evident in Eq.~\eqref{disp-iso}$_1$, the independently defined odd elastic stiffness components, $a_{1a}$ and $a_{1b}$, enter the longitudinal wave dispersion relationship through $A_\text{L}$. On the other hand, the transverse components of the elastic waves are not only fully degenerate (due to the absence of chiral coupling in centrosymmetric materials, $\mathbb{F}=\mathbb{G}=\mathbf{0}$), but they are also identical to their conservative counterparts. This equivalence is caused by the independence of the shear gradient stiffness, $A_\text{S}$, from the odd elasticity terms ($a_{1a}$ and $a_{1b}$). Crucially, while the active perturbations alter the longitudinal dispersion curve, the behavior of the odd elastic material remains indistinguishable from that of a passively stiffened conservative material. Because the odd elastic stiffness terms appear in the dispersion relationships only as their symmetric sum, $a_{1a}+a_{1b}$, wave propagation experiments can measure the total effective stiffening but cannot isolate the nonreciprocal asymmetry ($a_{1a} \neq a_{1b}$).
	
	\subsection{Wave Propagation in an Odd Elastic Hemitropic Material}
	\label{sec:res_hemitropic}
	
	In hemitropic materials, similar to isotropic materials, wave behavior is adequately characterized by analyzing propagation in one arbitrary direction. Accordingly, similar to isotropic materials, we focus on a plane wave propagating along the $x$-axis. Considering the chirality of the material, which results in surviving fifth-order elasticity tensors, Eq.~\eqref{Eq-Acoustic-Tensor} can be rewritten as
	\begin{equation}
		\Gamma_{ik} = C_{i1k1} k^2 + \mathrm{i} \left( F_{i1k11} - G_{i11k1} \right) k^3 + A_{i11k11} k^4 \, .
		\label{Q-x1-hemi}
	\end{equation}
	Utilizing the constitutive equations for hemitropic materials, Eq.~\eqref{PathA-Hemi-C}-\eqref{PathA-Hemi-G}, and substituting the above acoustic tensor into the eigenvalue equation of motion, Eq.~\eqref{eigval}, the condition for wave propagation in the $x$-direction in an odd elastic hemitropic material is derived as:
	\begin{equation}
		\begin{vmatrix}
			C_\text{L}k^2 + A_\text{L} k^4 - \rho \omega^2& 0 & 0 \\
			0 & C_\text{S} k^2 + A_\text{S} k^4 - \rho \omega^2 & \mathrm{i} (F_\text{S} + G_\text{S}) k^3 \\
			0 & -\mathrm{i} (F_\text{S} + G_\text{S}) k^3 & C_\text{S} k^2 + A_\text{S} k^4 - \rho \omega^2
		\end{vmatrix} 
		= 0,
		\label{dispersion-x1-hemit}
	\end{equation}
	where, in addition to the fourth- and sixth-order stiffness parameters given in Eqs.~\eqref{CSCL} and \eqref{ASAL}, $F_\text{S}$ and $G_\text{S}$ are the effective fifth-order chiral shear coupling stiffnesses:
	\begin{equation}
		\begin{aligned}
			F_\text{S} & \equiv F_{21311} = -F_{31211} = 2f_1 , \\
			G_\text{S} & \equiv G_{31121} = -G_{21131} = 2g_1  \, .
		\end{aligned}
	\end{equation}
	
	Comparing Eq.~\eqref{dispersion-x1-hemit} to its counterpart in the isotropic case, Eq.~\eqref{determ-x1-iso}, it is seen that chirality results only in the addition of off-diagonal components to the acoustic tensor. These off-diagonal components are complex conjugates of one another, ensuring that the acoustic tensor remains Hermitian. Accordingly, similar to the isotropic case, the system is guaranteed to possess purely real eigenvalues (as analytically established in Appendix \ref{app:hermiticity}). Consequently, despite the presence of active odd elastic moduli in both the fifth- and sixth-order stiffness tensors, wave propagation in hemitropic odd elastic materials remains elastodynamically conservative, preventing any complex frequencies or active energy exchange.

	Solving the eigenvalue problem above yields three distinct roots corresponding to one longitudinal and two transverse wave components
	
	\begin{equation}
		\begin{aligned}
			\omega_\text{L} &= \sqrt{\frac{C_\text{L}k^2 + A_\text{L} k^4}{\rho}} \,, \qquad & \text{Longitudinal component},\\
			\omega_\text{T} &=  \sqrt{\frac{C_\text{S} k^2 \pm (F_\text{S} + G_\text{S}) k^3 + A_\text{S} k^4}{\rho} } \,, \qquad & \text{Transverse components}.
		\end{aligned}
		\label{disp-hemi}
	\end{equation}
	The purely axial eigenvector corresponding to the first eigenvalue, $\mathbf{U}_\text{L} = (1, 0, 0)^{\mathsf{T}}$, confirms the purely longitudinal polarization of the first component, which is also fully decoupled from the chiral effects. 
	The two transverse frequencies, on the other hand, are coupled due to the chiral terms in the off-diagonal components of the acoustic tensor. The eigenvectors corresponding to these components are $\mathbf{U}_\text{T} = (0, \pm \mathrm{i}, 1)^{\mathsf{T}}$. The presence of the imaginary unit in the eigenvectors indicates a $90^\circ$ phase shift between the $y$ and $z$ components of wave oscillation, signifying that the transverse modes propagate as left-circularly and right-circularly polarized (LCP and RCP) shear waves \cite{portigal1968acoustical,auld1973acoustic}.

\begin{figure}[h]
		\centering
		\includegraphics[width=0.7\textwidth]{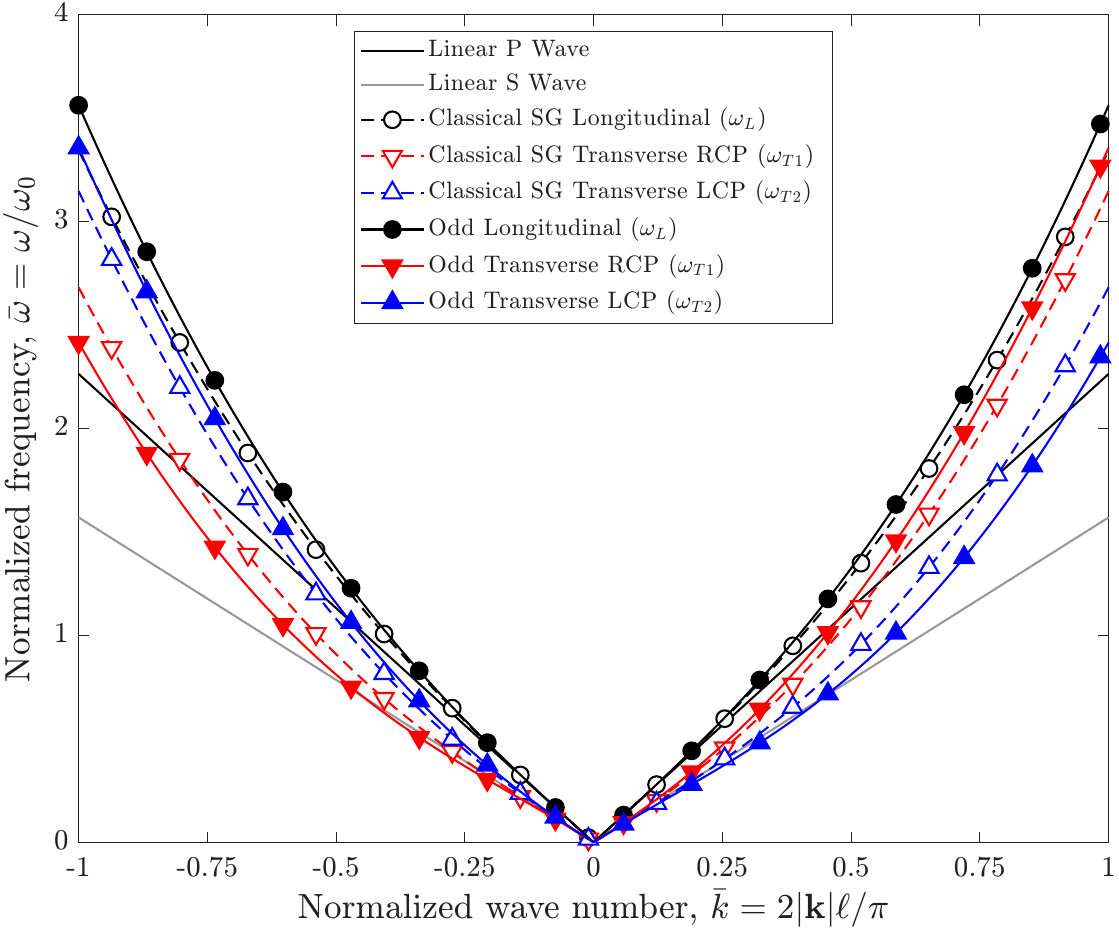}
		\caption{Normalized dispersion curves for a plane wave propagating in an odd elastic hemitropic continuum (solid lines with solid markers). The linear acoustic asymptotes (lines) and classical conservative strain-gradient limits (dashed lines with hollow markers) are provided for reference. We note the asymmetry of the transverse modes, caused by acoustical activity in chiral media, where a right-handed wave propagating forward is equivalent to a left-handed wave propagating backward ($\omega_{RCP}(k) = \omega_{LCP}(-k)$).}
		\label{fig:dispersion_hemi}
	\end{figure}

	For this numerical demonstration, we first introduce conservative chiral interactions in the form of hemitropic fifth-order stiffness tensors into the baseline isotropic material (Table \ref{tab:baseline_params}), yielding a passive hemitropic material. The fifth-order moduli need to scale with the first power of the length scale, $\ell$, from the first-gradient stiffness coefficients \cite{mindlin1965second, papanicolopulos2011chirality, pirmoradi2024anisotropic}. Accordingly, for the passive conservative case, we introduce chiral coupling moduli $f_1 =g_1= 0.2 \times 10^5$ N/m. Subsequently, to simulate an active continuum, we superimpose nonreciprocal interactions (\cite{scheibner2020odd, chen2021realization}). As demonstrated in Section \ref{sec:symmetries}, active nonreciprocal moduli, in the hemitropic (chiral) limit, manifest in the fifth-order stiffness tensors via two independent fifth-order tensors, $\mathbb{F}$ and $\mathbb{G}$. To achieve this odd elasticity, we perturb the $\mathbb{G}$ tensor only, and pick $g_1 = 3f_1$. The resulting normalized dispersion curves are plotted in Figure \ref{fig:dispersion_hemi} for both forward and backward propagation.

	Comparing Figure \ref{fig:dispersion_hemi} to Figure \ref{fig:dispersion_iso} reveals the distinct physical effects of the addition of chirality (via the addition of the fifth-order odd moduli). It is noted that the longitudinal component remains shifted relative to the conservative baseline due to the sixth-order odd parameters ($a_{1a} \neq a_{1b}$), similarly to the isotropic case. As opposed to the isotropic case, it is seen that revoking centrosymmetry and addition of chirality breaks the degeneracy of the shear waves and splits the transverse components. As demonstrated by the eigenvectors, this splitting corresponds to right-circularly and left-circularly polarized (RCP and LCP) transverse waves. It is also noted that the frequencies of these polarized components swap places when the direction of propagation changes, i.e., $\omega_\text{RCP}(k) = \omega_\text{LCP}(-k)$. This asymmetric behavior is a direct consequence of the wave polarizations, demonstrated by the eigenvectors of the RCP and LCP transverse waves. When the wave is propagating forward ($k > 0$), the RCP mode corresponds to $\mathbf{U}_\text{T} = (0, \mathrm{i}, 1)^{\mathsf{T}}$ and is governed by the $+k^3$ term in Eq.~\eqref{disp-hemi}$_2$, while the LCP mode corresponds to $\mathbf{U}_\text{T} = (0, -\mathrm{i}, 1)^{\mathsf{T}}$ and is governed by the $-k^3$ term. Reversing the propagation direction ($k \to -k$) flips the sign of the $k^3$ term. Therefore, the \textit{RCP wave propagating backward} is governed by the exact same dispersion relation as the \textit{LCP wave propagating forward}, and vice versa.
	
	Finally, while the presence of active nonreciprocal feedback alters both the longitudinal wave speed and the magnitude of the chiral splitting, the nonconservative nature of the material remains elastodynamically hidden. In classical conservative elasticity (shown by dashed lines with hollow markers in Figure \ref{fig:dispersion_hemi}), the magnitude of the transverse splitting is governed by $\pm 2F_\text{S} k^3$, whereas in the active medium, it is governed by $\pm(F_\text{S} + G_\text{S})k^3$. Because the fifth-order odd elastic parameters enter the dispersion relations only through this symmetric sum, the resulting behavior is indistinguishable from that of a passively stiffened hemitropic medium. Therefore, wave propagation experiments alone only measure the symmetric sums ($a_{1a}+a_{1b}$ and $f_1+g_1$) and cannot isolate the nonreciprocal differences ($a_{1a} \neq a_{1b}$ and $f_1 \neq g_1$). Consequently, dispersion tests cannot be used to distinguish a hemitropic odd elastic material from a conservative chiral medium.

	\subsection{Wave Propagation in an Odd Elastic Transversely Hemitropic Material}
	\label{sec:res_trans_hemitropic}
	
	As established in Appendix \ref{app:hermiticity} and numerically demonstrated in the previous examples, non-Hermitian elastodynamics, requires the elastic symmetry of the continuum to be reduced to allow the asymmetric components of the stiffness tensors to survive the wave vector contractions, Eq.~\eqref{Eq-Acoustic-Tensor}, into the acoustic tensor. Accordingly, we consider a transversely hemitropic material (Point Group $\infty 22$), with continuous rotational symmetry about the $z$-axis.
	To ensure the wave vector does not align with either the axis of continuous rotational symmetry, or the isotropy plane, we analyze an obliquely propagating plane wave with $\mathbf{k} = k(1, 0, 1)^{\mathsf{T}}/\sqrt{2}$.

	For this numerical demonstration, we perturb the isotropic granular assembly used in the previous examples to arrive at a realistic transversely hemitropic odd elastic material. To enforce transversely hemitropic odd elastic behavior, we break the 3D isotropy by stiffening the out-of-plane components and introduce active cross-couplings in the fourth-, fifth-, and sixth-order stiffness tensors.
	Physically, this system corresponds to a passive granular system, augmented with a tailored network of nonreciprocal active components (e.g., nonreciprocal sensor-actuator feedback loops \cite{brandenbourger2019non, chen2021realization}) with preferential alignment along the $z$-axis. As established in Section \ref{sec:TH} and Eqs.~\eqref{Cij-TH}-\eqref{F-TH}, this symmetry reduction results in $6$, $8$, and $31$ independent stiffness components for each of the fourth-, fifth- and sixth-order odd elastic stiffness tensors, respectively.
	For the fourth- and fifth-order stiffness tensors, $\mathbb{C}$, $\mathbb{F}$, and $\mathbb{G}$, we have stiffened the components corresponding to the direction of the axis of rotational symmetry ($z$-axis). For the sixth-order
	stiffness tensor, $\mathbb{A}$, the $31$ independent components, characterizing transversely hemitropic odd elastic materials are first generated for the hemitropic case studied in Section \ref{sec:res_hemitropic}. Afterwards, by perturbing all the
	derived parameters by a constant value, we break the hemitropic symmetry inherent in these parameters
	and derive $31$ independent transversely hemitropic stiffness parameters. For detailed description of the
	procedure and the resulting parameters used for the transveresely hemitropic odd elastic material, see Appendix \ref{app:THparams}.

	Substituting these parameters and the oblique wave vector into Eq.~\eqref{Eq-Acoustic-Tensor} yields a fully populated $3 \times 3$ acoustic tensor $\mathbf{\Gamma}$. For the acoustic tensor to be non-Hermitian, it is sufficient for either its real part ($\mathbf{\Gamma}^\text{R}$) to be asymmetric or its imaginary part ($\mathbf{\Gamma}^\text{I}$) to lack antisymmetry. In this transversely hemitropic odd elastic continuum, both conditions are met simultaneously: the real part of the acoustic tensor becomes asymmetric due to the anisotropy and asymmetry of $\mathbb{C}$ and $\mathbb{A}$, and the imaginary part loses its antisymmetry due to the anisotropy and independence of $\mathbb{F}$ and $\mathbb{G}$. 
	We note that while the first-gradient asymmetry, $\mathbb{C}\neq \mathbb{C}^{\mathsf{T}}$ (which, as was demonstrated in Section \ref{sec:symmetries}, also requires anisotropy), is sufficient to break Hermiticity \cite{scheibner2020odd, gao2022non, fruchart2023odd, ghatak2020observation, scheibner2020non, rosa2020dynamics, christensen2024perspective, zhou2026curved}, in strain-gradient odd elasticity, the higher-gradient tensors introduce additional non-Hermitian couplings. 
	
\begin{figure}[htpb]
		\centering
		\includegraphics[width=0.95\textwidth]{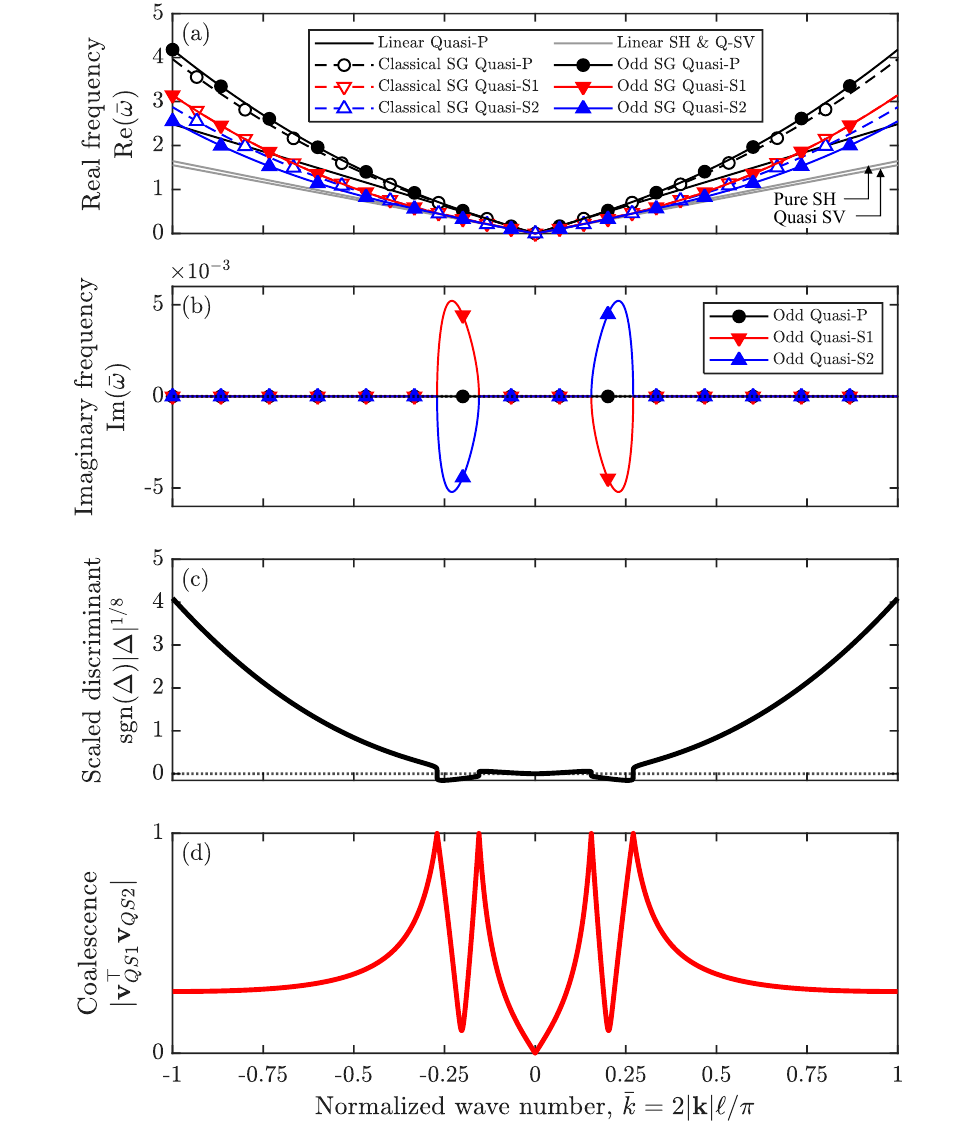}
		\caption{Dispersion analysis for an obliquely propagating wave in a transversely hemitropic odd elastic continuum: (a) the real part (solid lines with solid markers): first gradient asymptotes (lines) and classical conservative strain-gradient limits (dashed lines with hollow markers) are provided for reference, (b) the imaginary part: confirming the non-Hermitian nature of the acoustic tensor, (c) the scaled discriminant of the characteristic polynomial: showing the region with non-Hermitian behavior. (d) the coalescence of the eigenvectors of the two quasi-shear waves: showing exceptional points where eigenvectors (and eigenvalues) collide.}
		\label{fig:dispersion-TH}
	\end{figure}

	This interplay between first- and second-gradient nonreciprocity governs the dispersion relations, ultimately dictating the real/complex nature of frequency eigenvalues. In the example studied here, the acoustic tensor can be expanded as
	\begin{equation}
		\mathbf{\Gamma} = \mathbf{\Gamma}^\text{R} + \text{i} \mathbf{\Gamma}^\text{I}
		=
		\begin{bmatrix} \Gamma^\text{R}_{11} & 0 & \Gamma^\text{R}_{13} \\ 0 & \Gamma^\text{R}_{22} & 0 \\ \Gamma^\text{R}_{31} & 0 & \Gamma^\text{R}_{33} \end{bmatrix}
		+
		\text{i}\begin{bmatrix} 0 & \mathrm{i} \Gamma^\text{I}_{12} & 0 \\ \mathrm{i} \Gamma^\text{I}_{21} & 0 & \mathrm{i} \Gamma^\text{I}_{23} \\ 0 & \mathrm{i} \Gamma^\text{I}_{32} & 0 \end{bmatrix}
		=
		\begin{bmatrix} \Gamma^\text{R}_{11} & \mathrm{i} \Gamma^\text{I}_{12} & \Gamma^\text{R}_{13} \\ \mathrm{i} \Gamma^\text{I}_{21} & \Gamma^\text{R}_{22} & \mathrm{i} \Gamma^\text{I}_{23} \\ \Gamma^\text{R}_{31} & \mathrm{i} \Gamma^\text{I}_{32} & \Gamma^\text{R}_{33} \end{bmatrix} \, .
		\label{QrQi-TH}
	\end{equation}
	From this relationship it is seen that the characteristic equation that governs elastic wave propagation in these materials will be a polynomial with purely real coefficients. Accordingly, the real/complex nature of its resulting eigenvalues are fully governed by the real discriminant $\Delta$ of the polynomial. The results presented in Figure \ref{fig:dispersion-TH} show the emergence of complex frequency eigenvalues.

	Considering the complex nature of the wave frequencies, the dispersion curves are decomposed into their real and imaginary components. The evolution of the real part of the frequencies with varying wave number is shown in Figure \ref{fig:dispersion-TH}(a). The two conservative major-symmetric limits (first-gradient Cauchy limit and the standard conservative second-gradient limit) are also provided for reference.
	In the first-gradient limit, the imaginary part of the acoustic tensor vanishes and the acoustic tensor reduces to $\mathbf{\Gamma}^\text{R}$. We note from Eq.~\eqref{QrQi-TH} that $\mathbf{\Gamma}^\text{R}$ couples the $U_1$ and $U_3$ oscillation amplitudes while keeping $U_2$ decoupled. As a result, the eigenvector corresponding to $U_2$ is purely in the $y$-direction, representing a decoupled pure shear horizontal (SH) wave, while the eigenvectors corresponding to the coupled $U_1$ and $U_3$ components lie in the propagation plane ($xz$). Accordingly, the two coupled components include a quasi-primary (Quasi-P) wave and a quasi-shear vertical (Quasi-SV) wave (demonstrated in Figure \ref{fig:dispersion-TH}(a)).
	In the second-gradient models, both the conservative and odd elastic formulations couple all wave components, resulting in the splitting of all wave modes. We also note that, due to the even parity of $\mathbf{\Gamma}^\text{R}$ and the odd parity of $\mathbf{\Gamma}^\text{I}$ (demonstrated in Eq.~\eqref{QrQi-EvenOdd}) the real parts of the frequency eigenvalues are symmetric with respect to the wave number.

	For the odd elastic case, the imaginary parts of the frequency eigenvalues are reported in Figure \ref{fig:dispersion-TH}(b). Because the characteristic polynomial is of degree three with real coefficients, at least one eigenvalue must remain real, while the remaining two may form a complex conjugate pair. In this transversely hemitropic odd elastic material, we see that the Quasi-P component remains purely real at all wavelengths, while the two Quasi-S waves develop complex frequencies within a specific wave number band. Consistent with the odd parity of $\mathbf{\Gamma}^\text{I}$, these imaginary components are antisymmetric: a positive imaginary component in the forward direction corresponds to a negative imaginary component in the backward direction.

	The transition between purely real and complex conjugate frequencies is governed by the discriminant of the characteristic polynomial, $\Delta$. Figure \ref{fig:dispersion-TH}(c) shows the evolution of this discriminant with varying wave number. For increased readability, the signed eighth root of the discriminant, $\text{sgn}(\Delta)|\Delta|^{1/8}$, is presented to better demonstrate the regions with negative discriminant. The wave number regions in which frequency eigenvalues have imaginary components are bounded by the points at which the discriminant crosses zero, i.e., exceptional points (EPs) \cite{berry2004physics, heiss2012physics}. At these EPs, not only do the frequency eigenvalues degenerate, but their corresponding eigenvectors also coalesce \cite{mokhtari2020scattering,ashida2020non, fruchart2021non,miao2024negative,duan2025exceptional}. To explicitly visualize this collapse, the collinearity of the two normalized eigenvectors of the two quasi-shear components (Quasi-S1 and Quasi-S2) is presented via their Hermitian inner product, $|\bar{\mathbf{v}}_{QS1}^\mathsf{T} \mathbf{v}_{QS2}|$ in Figure \ref{fig:dispersion-TH}(d). These EPs mark the boundaries of the non-Hermitian instability band. At the onset of this region, the previously independent real QS modes collide and bifurcate into a complex conjugate pair. At the termination of this region, the complex conjugate modes coalesce once more and emerge as distinct, purely real propagating modes again \cite{scheibner2020non, zhou2020non}.

	Within this unstable domain, the elastodynamic response exhibits directional nonreciprocity. The emergence of complex conjugate frequency pairs indicates the simultaneous presence of amplifying (exponential energy growth) and attenuating (exponential energy decay) modes, in the positive and negative imaginary components, respectively.
	Because the exponential decay slows with time, while the growth compounds, growth of the amplifying mode dominates the response over time. Accordingly, the continuum experiences overall dynamic amplification.
	In any physical realization of such active metamaterials, this linear flutter instability \cite{mai2025anomalous} will ultimately be bounded by inevitable energy dissipation mechanisms.

	\section{Conclusion}
	\label{sec:conclusion}
	
	In this work, we have developed the second-gradient continuum theory (including anisotropy classification, balance laws and boundary conditions, followed by elastodynamics and wave propagation) for odd elasticity. By removing the assumption of the existence of a strain energy potential, the constitutive laws for second-gradient odd elasticity are formulated. Specifically, this expansion (i) relaxes the major symmetry requirement of the even-order stiffness tensors, i.e., $C_{ijkl} \neq C_{klij}$ and $A_{ijklmn} \neq A_{lmnijk}$, and (ii) relaxes the requirement that the two fifth-order coupling stiffness tensors must be transposes of one another, i.e., $F_{ijklm} \neq G_{klmij}$. By systematically applying elastic symmetries with respect to various rotation and reflection transformations (in chiral and achiral symmetry groups), we derived the exhaustive list of independent tensor components for all stiffness tensors, as well as the identities between nonzero components thereof. Further, utilizing the dynamic principle of virtual work, we derived the equilibrium equations and boundary conditions for these odd elastic second-gradient materials, followed by the resulting wave equations of motion. Finally, the equations of motion were used to solve elastic wave propagation within odd elastic second-gradient materials for specific elastic symmetry classes.
	
	The main conclusions of the paper are summarized below:
	
	\begin{enumerate}
		\item \textbf{Achiral 3D Odd First-Gradient Elasticity:} It is shown that in first-gradient elasticity (where stiffness is only characterized by the fourth-order stiffness tensor, $\mathbb{C}$), odd elastic stiffness components can be achieved in various chiral and achiral symmetry classes. The only symmetry conditions that \textit{do not} permit asymmetric fourth-order stiffness tensors are isotropy ($O(3)$), hemitropy ($SO(3)$), chiral cubic with 4-fold rotational symmetries ($432$), and its achiral counterpart ($m\bar{3}m$). Further, while in triclinic, monoclinic, orthorhombic, and orthotropic cases all off-diagonal components can be asymmetric, in the transversely hemitropic/isotropic ($\infty22$ and $\infty/mmm$) and cubic (both chiral and achiral) cases ($23$ and $m\bar{3}$), odd elasticity only manifests via a single odd-elastic parameter. In the transversely isotropic/hemitropic case, the application of continuous rotation about one axis makes the in-plane coupling components symmetric, Eq.~\eqref{PathA-Transverse-C}$_3$, but those coupling the axis of symmetry with the other axes remain odd. On the other hand, in the cubic case, a single odd modulus decouples all forward-cyclic couplings ($C_{12} = C_{23} = C_{31}$) from the backward-cyclic couplings ($C_{32} = C_{21} = C_{13}$), Eq.~\eqref{PathA-Cubic-C}$_{2,3}$.
		
		\item \textbf{Survival of Strain-Gradient Odd Elasticity in 3D Isotropy:} While continuous 3D rotation symmetry, $SO(3)$, prevents major asymmetry in the fourth-order tensor $\mathbb{C}$, odd elasticity is shown to survive in second-gradient terms. The sixth-order tensor $\mathbb{A}$ demonstrates nonconservative odd elastic behavior, even in the isotropic, $O(3)$, limit, via the split of the first higher-order parameter $a_1$ into $a_{1a} \neq a_{1b}$. In different levels of anisotropy, the sixth-order stiffness tensor retains additional asymmetric terms. Similar to the fourth-order stiffness tensor, in triclinic, monoclinic, orthorhombic, and orthotropic cases, all off-diagonal components of $\mathbb{A}$ can show asymmetric components in odd elasticity. On the other hand, in the transversely hemitropic/isotropic cases, the application of continuous rotation about one axis enforces major symmetry on three components in the in-plane direction, but allows $10$ asymmetric out-of-plane off-diagonal components, Eqs.~\eqref{PathA-Transverse-Forced-Major}, \eqref{PathA-Transverse-A-odd}. In the cubic case, 11 asymmetric off-diagonal components include one component that decouples forward and backward cyclic orbits, Eq.~\eqref{PathA-Cubic-A}$_{2,3}$, and $10$ additional asymmetric components, Eq.~\eqref{PathA-Cubic-A}$_4$. Moreover, even in the hemitropic case, the fifth-order stiffness tensors, $\mathbb{F}$ and $\mathbb{G}$, also demonstrate odd elasticity via independent components $f_1 \neq g_1$.
		
		\item \textbf{Dynamical Invisibility of Odd Elasticity in Highly Symmetric Media:} Despite the survival of odd elastic stiffness parameters in isotropic and hemitropic symmetry classes, their elastic wave dispersion is shown to not manifest the nonreciprocal nature of these effects. In effect, the acoustic tensors for materials with $O(3)$ and $SO(3)$ symmetries contain \textit{the symmetric sums} of the (odd) elastic stiffness coefficients only (i.e., $a_{1a}+a_{1b}$ and, in the case of chiral hemitropic materials, also $f_1+g_1$). Neither of these combinations can demonstrate the effect of the split between the two parameters caused by odd elasticity. Therefore, elastic wave dispersion in odd elastic isotropic and hemitropic materials is indistinguishable from that in conservative passive elastic media. Accordingly, while elastic wave propagation experiments can measure the overall shift in effective stiffness caused by activity, they cannot be used to isolate the individual nonreciprocal parameters ($a_{1a} \neq a_{1b}$ and $f_1 \neq g_1$) or quantify the degree of odd elasticity in highly symmetric media.

		\item \textbf{Non-Hermitian Behavior and Exceptional Points:} Non-Hermitian elastodynamics, characterized by complex frequency eigenvalues and active energy exchange (gain/loss), is shown to require not only nonconservative odd elasticity, but also a lack of complete rotational elastic symmetry. In the example of a transversely hemitropic material, the asymmetric off-diagonal components of $\mathbb{C}$ and $\mathbb{A}$ survive into the real part of the acoustic tensor, breaking its symmetry. Concurrently, the independence of the fifth-order coupling tensors ($\mathbb{F} \neq \mathbb{G}^\mathsf{T}$) breaks the antisymmetry of the imaginary part. The transition between conservative and active wave propagation is marked by exceptional points, where both eigenvalues and eigenvectors of two of the wave components coalesce. Within the unstable wavelengths (bounded by these EPs) where frequency becomes complex, the positive and negative imaginary components signify exponential growth and decay in energy, making wave propagation nonconservative. 
	\end{enumerate}
	
	It is emphasized that the numerical elastodynamic examples studied here, rather than reflecting a specific design of an active metamaterial, were constructed by superimposing active nonreciprocal perturbations onto a conservative granular baseline. By maintaining physically realistic stiffness, density, and length scales, these calculations demonstrate the fundamental dynamic consequences of higher-gradient odd elasticity and different levels of anisotropy. Accordingly, the present work establishes the exact symmetry and gradient-order requirements to achieve non-Hermitian dispersion. Also, the exhaustive classification of higher-order tensors provides the necessary theoretical foundation for the future inverse-design, calibration, and experimental realization of active, nonreciprocal mechanical metamaterials. In this direction. expanding multiscale homogenization frameworks to derive the connection between the component-scale active nonconservative properties on one hand, and the macroscopic strain-gradient odd elastic stiffness tensors on the other (at different levels of anisotropy), is an important topic of further research.

	\section*{Data and Code Availability}
	The data and codes that support the findings of this study can be directly reproduced from the information provided and are available upon reasonable request to the author.
	
	\section*{Declaration of Competing Interest}
	The author declares that there are no competing interests that could have appeared to influence the work.

	\begin{appendices}
		
		\counterwithin{equation}{section}
		\counterwithin{figure}{section}
		\counterwithin{table}{section}
		
		\renewcommand{\theequation}{\thesection.\arabic{equation}}
		\renewcommand{\thefigure}{\thesection.\arabic{figure}}
		\renewcommand{\thetable}{\thesection.\arabic{table}}
		
		\section{Voigt Mapping for Higher-Order Tensors}
		\label{app:voigt}
		
		The correspondence between the indices of the second- and third-order stress and strain tensors and their Voigt representations ($\alpha, \beta, \gamma$) are provided in Tables \ref{ij-alpha} and \ref{ijk-alpha}.
		
		\begin{table}[h]
			\centering
			\begin{tabular}{|l|c|c|c|c|c|c|}
				\hline
				Tensor indices $ij$ &$11$&$22$&$33$&$23$&$13$&$12$ \\  
				Voigt index $\alpha$ &$1$&$2$&$3$&$4$&$5$&$6$ \\
				\hline
			\end{tabular}
			\caption{Index mapping for second-order tensors to Voigt notation.}
			\label{ij-alpha}
		\end{table}
		
		\begin{table}[h]
			\centering
			\begin{tabular}{|l|c|c|c|c|c|c|c|c|c|}
				\hline
				Tensor indices $ijk$ &$111$&$122$&$133$&$123$&$113$&$112$\\
				Voigt index $\alpha$ &$01$&$02$&$03$&$04$&$05$&$06$\\
				\hline
				Tensor indices $ijk$ &$211$&$222$&$233$&$223$&$213$&$212$\\
				Voigt index $\alpha$ &$07$&$08$&$09$&$10$&$11$&$12$\\
				\hline
				Tensor indices $ijk$ &$311$&$322$&$333$&$323$&$313$&$312$\\
				Voigt index $\alpha$ &$13$&$14$&$15$&$16$&$17$&$18$\\
				\hline
			\end{tabular}
			\caption{Index mapping for third-order tensors to Voigt notation.}
			\label{ijk-alpha}
		\end{table}

		\section{Explicit Matrix Representations of Odd Elastic Tensors}
		\label{app:matrices}
		
		This appendix provides the exhaustive matrix representations of the odd elastic stiffness tensors ($\mathbb{C}$, $\mathbb{F}$, and $\mathbb{A}$) in Voigt notation for all chiral and achiral symmetry classes discussed in the paper. In each class, the matrix representation of the other fifth-order tensor, $\mathbb{G}$, can be obtained trivially by transposing $\mathbb{F}$ and replacing the moduli $F_{\alpha\,\beta}$ with independent moduli $G_{\beta\,\alpha}$. 
		For completeness, we also present the number of independent components of odd stiffness tensors for all fundamental elastic symmetries in Section \ref{app:expanded_classification}.

		\subsection{Triclinic Classes (Point Group 1)}
		In the triclinic class, the material lacks any rotational or reflection symmetry. Consequently, all components of the stiffness tensors are independent and nonzero. The matrices $\mathbf{C}_{6 \times 6}$, $\mathbf{F}_{6 \times 18}$, $\mathbf{G}_{18 \times 6}$, and $\mathbf{A}_{18 \times 18}$ are all fully populated, with 36, 108, 108, and 324 independent components, respectively.
		
		\subsection{Monoclinic Classes (Point Groups 2 and $m$)} \label{app:sec:mono}
		The monoclinic classes possess a single axis of two-fold rotation (Point Group 2) or a single plane of reflection (Point Group $m$).
		To systematically evaluate the monoclinic symmetry classes, the Voigt indices are partitioned into disjoint sets based on the parity of index 1 ($p_1$). In classical tensor notation, these sets are derived as:
		\begin{equation}
			\begin{aligned}
				\mathcal{V}_0 &= \{11, 22, 33, 23\} \quad &&(4 \text{ components with } p_1=0), \\
				\mathcal{V}_1 &= \{13, 12\} \quad &&(2 \text{ components with } p_1=1), \\
				\mathcal{K}_0 &= \{113, 112, 211, 222, 233, 223, 311, 322, 333, 323\} \quad &&(10 \text{ components with } p_1=0), \\
				\mathcal{K}_1 &= \{111, 122, 133, 123, 213, 212, 313, 312\} \quad &&(8 \text{ components with } p_1=1).
			\end{aligned}
			\label{parity-tensor}
		\end{equation}
		Using the mappings from Appendix \ref{app:voigt}, the partitions are reformulated into Voigt notation as:
		\begin{equation}
			\begin{aligned}
				\mathcal{V}_0 &= \{1, 2, 3, 4\}, \quad &\mathcal{V}_1 &= \{5, 6\}, \\
				\mathcal{K}_0 &= \{05, 06, 07, 08, 09, 10, 13, 14, 15, 16\}, \quad &\mathcal{K}_1 &= \{01, 02, 03, 04, 11, 12, 17, 18\}.
			\end{aligned}
			\label{parity-voigt}
		\end{equation}
		The even-order tensors ($\mathbf{C}$ and $\mathbf{A}$) are identical for both classes. In the case where the $x$-axis is the axis of rotation (or $yz$-plane is the plane of reflection), the fourth-order tensor $\mathbf{C}$ (with 20 independent parameters) reads:
		\begin{equation}
			\mathbf{C}_{\text{mono}} = \begin{bmatrix}  
				C_{11} & C_{12} & C_{13} & C_{14} & 0 & 0\\
				C_{21} & C_{22} & C_{23} & C_{24} & 0 & 0\\ 
				C_{31} & C_{32} & C_{33} & C_{34} & 0 & 0\\ 
				C_{41} & C_{42} & C_{43} & C_{44} & 0 & 0\\
				0 & 0 & 0 & 0 & C_{55} & C_{56}\\ 
				0 & 0 & 0 & 0 & C_{65} & C_{66}
			\end{bmatrix} \, .
			\label{Cij-mono}
		\end{equation} 
		Similarly, the sixth-order tensor $\mathbb{A}$ (with 164 independent components) reads:
		\setcounter{MaxMatrixCols}{18}
		\begin{equation}
			\resizebox{\textwidth}{!}{
				$ \mathbf{A}_{\text{mono}} = \begin{bmatrix}
					A_{01\,01} & A_{01\,02} & A_{01\,03} & A_{01\,04} & 0 & 0 & 0 & 0 & 0 & 0 & A_{01\,11} & A_{01\,12} & 0 & 0 & 0 & 0 & A_{01\,17} & A_{01\,18} \\
					A_{02\,01} & A_{02\,02} & A_{02\,03} & A_{02\,04} & 0 & 0 & 0 & 0 & 0 & 0 & A_{02\,11} & A_{02\,12} & 0 & 0 & 0 & 0 & A_{02\,17} & A_{02\,18} \\
					A_{03\,01} & A_{03\,02} & A_{03\,03} & A_{03\,04} & 0 & 0 & 0 & 0 & 0 & 0 & A_{03\,11} & A_{03\,12} & 0 & 0 & 0 & 0 & A_{03\,17} & A_{03\,18} \\
					A_{04\,01} & A_{04\,02} & A_{04\,03} & A_{04\,04} & 0 & 0 & 0 & 0 & 0 & 0 & A_{04\,11} & A_{04\,12} & 0 & 0 & 0 & 0 & A_{04\,17} & A_{04\,18} \\
					0 & 0 & 0 & 0 & A_{05\,05} & A_{05\,06} & A_{05\,07} & A_{05\,08} & A_{05\,09} & A_{05\,10} & 0 & 0 & A_{05\,13} & A_{05\,14} & A_{05\,15} & A_{05\,16} & 0 & 0 \\
					0 & 0 & 0 & 0 & A_{06\,05} & A_{06\,06} & A_{06\,07} & A_{06\,08} & A_{06\,09} & A_{06\,10} & 0 & 0 & A_{06\,13} & A_{06\,14} & A_{06\,15} & A_{06\,16} & 0 & 0 \\
					0 & 0 & 0 & 0 & A_{07\,05} & A_{07\,06} & A_{07\,07} & A_{07\,08} & A_{07\,09} & A_{07\,10} & 0 & 0 & A_{07\,13} & A_{07\,14} & A_{07\,15} & A_{07\,16} & 0 & 0 \\
					0 & 0 & 0 & 0 & A_{08\,05} & A_{08\,06} & A_{08\,07} & A_{08\,08} & A_{08\,09} & A_{08\,10} & 0 & 0 & A_{08\,13} & A_{08\,14} & A_{08\,15} & A_{08\,16} & 0 & 0 \\
					0 & 0 & 0 & 0 & A_{09\,05} & A_{09\,06} & A_{09\,07} & A_{09\,08} & A_{09\,09} & A_{09\,10} & 0 & 0 & A_{09\,13} & A_{09\,14} & A_{09\,15} & A_{09\,16} & 0 & 0 \\
					0 & 0 & 0 & 0 & A_{10\,05} & A_{10\,06} & A_{10\,07} & A_{10\,08} & A_{10\,09} & A_{10\,10} & 0 & 0 & A_{10\,13} & A_{10\,14} & A_{10\,15} & A_{10\,16} & 0 & 0 \\
					A_{11\,01} & A_{11\,02} & A_{11\,03} & A_{11\,04} & 0 & 0 & 0 & 0 & 0 & 0 & A_{11\,11} & A_{11\,12} & 0 & 0 & 0 & 0 & A_{11\,17} & A_{11\,18} \\
					A_{12\,01} & A_{12\,02} & A_{12\,03} & A_{12\,04} & 0 & 0 & 0 & 0 & 0 & 0 & A_{12\,11} & A_{12\,12} & 0 & 0 & 0 & 0 & A_{12\,17} & A_{12\,18} \\
					0 & 0 & 0 & 0 & A_{13\,05} & A_{13\,06} & A_{13\,07} & A_{13\,08} & A_{13\,09} & A_{13\,10} & 0 & 0 & A_{13\,13} & A_{13\,14} & A_{13\,15} & A_{13\,16} & 0 & 0 \\
					0 & 0 & 0 & 0 & A_{14\,05} & A_{14\,06} & A_{14\,07} & A_{14\,08} & A_{14\,09} & A_{14\,10} & 0 & 0 & A_{14\,13} & A_{14\,14} & A_{14\,15} & A_{14\,16} & 0 & 0 \\
					0 & 0 & 0 & 0 & A_{15\,05} & A_{15\,06} & A_{15\,07} & A_{15\,08} & A_{15\,09} & A_{15\,10} & 0 & 0 & A_{15\,13} & A_{15\,14} & A_{15\,15} & A_{15\,16} & 0 & 0 \\
					0 & 0 & 0 & 0 & A_{16\,05} & A_{16\,06} & A_{16\,07} & A_{16\,08} & A_{16\,09} & A_{16\,10} & 0 & 0 & A_{16\,13} & A_{16\,14} & A_{16\,15} & A_{16\,16} & 0 & 0 \\
					A_{17\,01} & A_{17\,02} & A_{17\,03} & A_{17\,04} & 0 & 0 & 0 & 0 & 0 & 0 & A_{17\,11} & A_{17\,12} & 0 & 0 & 0 & 0 & A_{17\,17} & A_{17\,18} \\
					A_{18\,01} & A_{18\,02} & A_{18\,03} & A_{18\,04} & 0 & 0 & 0 & 0 & 0 & 0 & A_{18\,11} & A_{18\,12} & 0 & 0 & 0 & 0 & A_{18\,17} & A_{18\,18}
				\end{bmatrix} \, . $}
			\label{A-mono}
		\end{equation}
		
		For the fifth-order tensor $\mathbb{F}$ the nonzero components depend on whether the material is chiral or achiral. In the chiral monoclinic case (Point Group $2$),  $\mathbb{F}$ has 52 independent constants and reads:
		\begin{equation}
			\resizebox{\textwidth}{!}{
				$ \mathbf{F}_{2} = \begin{bmatrix}
					F_{1\,01} & F_{1\,02} & F_{1\,03} & F_{1\,04} & 0 & 0 & 0 & 0 & 0 & 0 & F_{1\,11} & F_{1\,12} & 0 & 0 & 0 & 0 & F_{1\,17} & F_{1\,18} \\
					F_{2\,01} & F_{2\,02} & F_{2\,03} & F_{2\,04} & 0 & 0 & 0 & 0 & 0 & 0 & F_{2\,11} & F_{2\,12} & 0 & 0 & 0 & 0 & F_{2\,17} & F_{2\,18} \\
					F_{3\,01} & F_{3\,02} & F_{3\,03} & F_{3\,04} & 0 & 0 & 0 & 0 & 0 & 0 & F_{3\,11} & F_{3\,12} & 0 & 0 & 0 & 0 & F_{3\,17} & F_{3\,18} \\
					F_{4\,01} & F_{4\,02} & F_{4\,03} & F_{4\,04} & 0 & 0 & 0 & 0 & 0 & 0 & F_{4\,11} & F_{4\,12} & 0 & 0 & 0 & 0 & F_{4\,17} & F_{4\,18} \\
					0 & 0 & 0 & 0 & F_{5\,05} & F_{5\,06} & F_{5\,07} & F_{5\,08} & F_{5\,09} & F_{5\,10} & 0 & 0 & F_{5\,13} & F_{5\,14} & F_{5\,15} & F_{5\,16} & 0 & 0 \\
					0 & 0 & 0 & 0 & F_{6\,05} & F_{6\,06} & F_{6\,07} & F_{6\,08} & F_{6\,09} & F_{6\,10} & 0 & 0 & F_{6\,13} & F_{6\,14} & F_{6\,15} & F_{6\,16} & 0 & 0
				\end{bmatrix} \, . $}
			\label{F-mono-chiral}
		\end{equation}
		On the other hand, for the achiral monoclinic case (Point Group $m$), the tensor has nonzero values in the 56 complementary components and reads:
		\begin{equation}
			\resizebox{\textwidth}{!}{
				$ \mathbf{F}_{m} = \begin{bmatrix}
					0 & 0 & 0 & 0 & F_{1\,05} & F_{1\,06} & F_{1\,07} & F_{1\,08} & F_{1\,09} & F_{1\,10} & 0 & 0 & F_{1\,13} & F_{1\,14} & F_{1\,15} & F_{1\,16} & 0 & 0 \\
					0 & 0 & 0 & 0 & F_{2\,05} & F_{2\,06} & F_{2\,07} & F_{2\,08} & F_{2\,09} & F_{2\,10} & 0 & 0 & F_{2\,13} & F_{2\,14} & F_{2\,15} & F_{2\,16} & 0 & 0 \\
					0 & 0 & 0 & 0 & F_{3\,05} & F_{3\,06} & F_{3\,07} & F_{3\,08} & F_{3\,09} & F_{3\,10} & 0 & 0 & F_{3\,13} & F_{3\,14} & F_{3\,15} & F_{3\,16} & 0 & 0 \\
					0 & 0 & 0 & 0 & F_{4\,05} & F_{4\,06} & F_{4\,07} & F_{4\,08} & F_{4\,09} & F_{4\,10} & 0 & 0 & F_{4\,13} & F_{4\,14} & F_{4\,15} & F_{4\,16} & 0 & 0 \\
					F_{5\,01} & F_{5\,02} & F_{5\,03} & F_{5\,04} & 0 & 0 & 0 & 0 & 0 & 0 & F_{5\,11} & F_{5\,12} & 0 & 0 & 0 & 0 & F_{5\,17} & F_{5\,18} \\
					F_{6\,01} & F_{6\,02} & F_{6\,03} & F_{6\,04} & 0 & 0 & 0 & 0 & 0 & 0 & F_{6\,11} & F_{6\,12} & 0 & 0 & 0 & 0 & F_{6\,17} & F_{6\,18}
				\end{bmatrix} \, . $}
			\label{F-mono-achiral}
		\end{equation}

		\subsection{Orthorhombic Classes (Point Groups 222, $mm2$, and $mmm$)}
		The orthorhombic classes are characterized by three mutually orthogonal directions of symmetry, which stem from different elastic symmetries in three point groups: three two-fold axes (222), two mirror planes and one two-fold axis ($mm2$), and three mutually orthogonal mirror planes ($mmm$). 
		For these classes, the components of the second- and third-order tensors are partitioned based on the parity triplet $\mathbf{p} = p_1 p_2 p_3$, as detailed in Table \ref{tab:Orthorho-Sets}.
		
		\begin{table}[h]
			\centering
			\renewcommand{\arraystretch}{1.5}
			\begin{tabular}{c l c | c l c}
				\hline
				\multicolumn{3}{c|}{\textbf{Second-order tensors}} & \multicolumn{3}{c}{\textbf{Third-order tensors}} \\
				\hline
				\hline
				Parity $\mathbf{p}$ & $\mathcal{V}_\mathbf{p}$ (Tensor) & $\mathcal{V}_\mathbf{p}$ (Voigt) & Parity $\mathbf{p}$ & $\mathcal{K}_\mathbf{p}$ (Tensor) & $\mathcal{K}_\mathbf{p}$ (Voigt) \\
				\hline
				$000$ & $\{11, 22, 33\}$ & $\{1, 2, 3\}$ & $111$ & $\{123, 213, 312\}$           & $\{04, 11, 18\}$  \\
				$011$ & $\{23\}$         & $\{4\}$       & $100$ & $\{111, 122, 133, 212, 313\}$ & $\{01, 02, 03, 12, 17\}$ \\
				$101$ & $\{13\}$         & $\{5\}$       & $010$ & $\{112, 211, 222, 233, 323\}$ & $\{06, 07, 08, 09, 16\}$ \\
				$110$ & $\{12\}$         & $\{6\}$       & $001$ & $\{113, 223, 311, 322, 333\}$ & $\{05, 10, 13, 14, 15\}$\\
				\hline
			\end{tabular}
			\caption{Partitioning of the classical strain ($\mathcal{V}_\mathbf{p}$) and strain gradient ($\mathcal{K}_\mathbf{p}$) components into sets based on their parity triplet $\mathbf{p}=p_1 p_2 p_3$.}
			\label{tab:Orthorho-Sets}
		\end{table}

		Because even-order tensors are invariant under central inversion, the constraints for $\mathbf{C}$ and $\mathbf{A}$ are identical across all three groups. The fourth-order tensor $\mathbf{C}$ (with 12 independent parameters) is given by:
		\begin{equation}
			\mathbf{C}_{\text{ortho}} = \begin{bmatrix}  
				C_{11} & C_{12} & C_{13} & 0 & 0 & 0\\
				C_{21} & C_{22} & C_{23} & 0 & 0 & 0\\ 
				C_{31} & C_{32} & C_{33} & 0 & 0 & 0\\ 
				0 & 0 & 0 & C_{44} & 0 & 0\\
				0 & 0 & 0 & 0 & C_{55} & 0\\ 
				0 & 0 & 0 & 0 & 0 & C_{66}
			\end{bmatrix} \, .
			\label{Cij-ortho}
		\end{equation} 
		Similarly, the sixth-order tensor $\mathbb{A}$ (with 84 independent components) reads:
		\setcounter{MaxMatrixCols}{18}
		\begin{equation}
			\resizebox{\textwidth}{!}{
				$ \mathbf{A}_{\text{ortho}} = \begin{bmatrix}
					A_{01\,01} & A_{01\,02} & A_{01\,03} & 0 & 0 & 0 & 0 & 0 & 0 & 0 & 0 & A_{01\,12} & 0 & 0 & 0 & 0 & A_{01\,17} & 0 \\
					A_{02\,01} & A_{02\,02} & A_{02\,03} & 0 & 0 & 0 & 0 & 0 & 0 & 0 & 0 & A_{02\,12} & 0 & 0 & 0 & 0 & A_{02\,17} & 0 \\
					A_{03\,01} & A_{03\,02} & A_{03\,03} & 0 & 0 & 0 & 0 & 0 & 0 & 0 & 0 & A_{03\,12} & 0 & 0 & 0 & 0 & A_{03\,17} & 0 \\
					0 & 0 & 0 & A_{04\,04} & 0 & 0 & 0 & 0 & 0 & 0 & A_{04\,11} & 0 & 0 & 0 & 0 & 0 & 0 & A_{04\,18} \\
					0 & 0 & 0 & 0 & A_{05\,05} & 0 & 0 & 0 & 0 & A_{05\,10} & 0 & 0 & A_{05\,13} & A_{05\,14} & A_{05\,15} & 0 & 0 & 0 \\
					0 & 0 & 0 & 0 & 0 & A_{06\,06} & A_{06\,07} & A_{06\,08} & A_{06\,09} & 0 & 0 & 0 & 0 & 0 & 0 & A_{06\,16} & 0 & 0 \\
					0 & 0 & 0 & 0 & 0 & A_{07\,06} & A_{07\,07} & A_{07\,08} & A_{07\,09} & 0 & 0 & 0 & 0 & 0 & 0 & A_{07\,16} & 0 & 0 \\
					0 & 0 & 0 & 0 & 0 & A_{08\,06} & A_{08\,07} & A_{08\,08} & A_{08\,09} & 0 & 0 & 0 & 0 & 0 & 0 & A_{08\,16} & 0 & 0 \\
					0 & 0 & 0 & 0 & 0 & A_{09\,06} & A_{09\,07} & A_{09\,08} & A_{09\,09} & 0 & 0 & 0 & 0 & 0 & 0 & A_{09\,16} & 0 & 0 \\
					0 & 0 & 0 & 0 & A_{10\,05} & 0 & 0 & 0 & 0 & A_{10\,10} & 0 & 0 & A_{10\,13} & A_{10\,14} & A_{10\,15} & 0 & 0 & 0 \\
					0 & 0 & 0 & A_{11\,04} & 0 & 0 & 0 & 0 & 0 & 0 & A_{11\,11} & 0 & 0 & 0 & 0 & 0 & 0 & A_{11\,18} \\
					A_{12\,01} & A_{12\,02} & A_{12\,03} & 0 & 0 & 0 & 0 & 0 & 0 & 0 & 0 & A_{12\,12} & 0 & 0 & 0 & 0 & A_{12\,17} & 0 \\
					0 & 0 & 0 & 0 & A_{13\,05} & 0 & 0 & 0 & 0 & A_{13\,10} & 0 & 0 & A_{13\,13} & A_{13\,14} & A_{13\,15} & 0 & 0 & 0 \\
					0 & 0 & 0 & 0 & A_{14\,05} & 0 & 0 & 0 & 0 & A_{14\,10} & 0 & 0 & A_{14\,13} & A_{14\,14} & A_{14\,15} & 0 & 0 & 0 \\
					0 & 0 & 0 & 0 & A_{15\,05} & 0 & 0 & 0 & 0 & A_{15\,10} & 0 & 0 & A_{15\,13} & A_{15\,14} & A_{15\,15} & 0 & 0 & 0 \\
					0 & 0 & 0 & 0 & 0 & A_{16\,06} & A_{16\,07} & A_{16\,08} & A_{16\,09} & 0 & 0 & 0 & 0 & 0 & 0 & A_{16\,16} & 0 & 0 \\
					A_{17\,01} & A_{17\,02} & A_{17\,03} & 0 & 0 & 0 & 0 & 0 & 0 & 0 & 0 & A_{17\,12} & 0 & 0 & 0 & 0 & A_{17\,17} & 0 \\
					0 & 0 & 0 & A_{18\,04} & 0 & 0 & 0 & 0 & 0 & 0 & A_{18\,11} & 0 & 0 & 0 & 0 & 0 & 0 & A_{18\,18}
				\end{bmatrix} \, . $}
			\label{A-ortho}
		\end{equation}
		For the fifth-order tensors $\mathbb{F}$, the specific symmetry mechanism governs the nonzero components of the stiffness tensor. In the chiral orthorhombic class (Point Group $222$) $\mathbb{F}$ read:
		\begin{equation}
			\resizebox{\textwidth}{!}{
				$ \mathbf{F}_{222} = \begin{bmatrix}
					0 & 0 & 0 & F_{1\,04} & 0 & 0 & 0 & 0 & 0 & 0 & F_{1\,11} & 0 & 0 & 0 & 0 & 0 & 0 & F_{1\,18} \\
					0 & 0 & 0 & F_{2\,04} & 0 & 0 & 0 & 0 & 0 & 0 & F_{2\,11} & 0 & 0 & 0 & 0 & 0 & 0 & F_{2\,18} \\
					0 & 0 & 0 & F_{3\,04} & 0 & 0 & 0 & 0 & 0 & 0 & F_{3\,11} & 0 & 0 & 0 & 0 & 0 & 0 & F_{3\,18} \\
					F_{4\,01} & F_{4\,02} & F_{4\,03} & 0 & 0 & 0 & 0 & 0 & 0 & 0 & 0 & F_{4\,12} & 0 & 0 & 0 & 0 & F_{4\,17} & 0 \\
					0 & 0 & 0 & 0 & 0 & F_{5\,06} & F_{5\,07} & F_{5\,08} & F_{5\,09} & 0 & 0 & 0 & 0 & 0 & 0 & F_{5\,16} & 0 & 0 \\
					0 & 0 & 0 & 0 & F_{6\,05} & 0 & 0 & 0 & 0 & F_{6\,10} & 0 & 0 & F_{6\,13} & F_{6\,14} & F_{6\,15} & 0 & 0 & 0
				\end{bmatrix} \,  $},
			\label{F-222}
		\end{equation}
		while for the achiral orthorhombic class (Point Group $mm2$), it reads:
		\begin{equation}
			\resizebox{\textwidth}{!}{
				$ \mathbf{F}_{mm2} = \begin{bmatrix}
					0 & 0 & 0 & 0 & F_{1\,05} & 0 & 0 & 0 & 0 & F_{1\,10} & 0 & 0 & F_{1\,13} & F_{1\,14} & F_{1\,15} & 0 & 0 & 0 \\
					0 & 0 & 0 & 0 & F_{2\,05} & 0 & 0 & 0 & 0 & F_{2\,10} & 0 & 0 & F_{2\,13} & F_{2\,14} & F_{2\,15} & 0 & 0 & 0 \\
					0 & 0 & 0 & 0 & F_{3\,05} & 0 & 0 & 0 & 0 & F_{3\,10} & 0 & 0 & F_{3\,13} & F_{3\,14} & F_{3\,15} & 0 & 0 & 0 \\
					0 & 0 & 0 & 0 & 0 & F_{4\,06} & F_{4\,07} & F_{4\,08} & F_{4\,09} & 0 & 0 & 0 & 0 & 0 & 0 & F_{4\,16} & 0 & 0 \\
					F_{5\,01} & F_{5\,02} & F_{5\,03} & 0 & 0 & 0 & 0 & 0 & 0 & 0 & 0 & F_{5\,12} & 0 & 0 & 0 & 0 & F_{5\,17} & 0 \\
					0 & 0 & 0 & F_{6\,04} & 0 & 0 & 0 & 0 & 0 & 0 & F_{6\,11} & 0 & 0 & 0 & 0 & 0 & 0 & F_{6\,18}
				\end{bmatrix} \, . $}
			\label{F-mm2}
		\end{equation}
		Finally, for the centrosymmetric orthotropic class (Point Group $mmm$), the presence of a center of inversion ensures that odd-order tensors vanish entirely:
		\begin{equation}
			\mathbf{F}_{mmm} = \mathbf{0} \, .
			\label{F-mmm}
		\end{equation}

		\subsection{Transversely Hemitropic ($\infty 22$) and Transversely Isotropic ($\infty/mmm$)}
		The transversely hemitropic and transversely isotropic classes are generated by adding an axis of infinite rotation symmetry to the chiral orthorhombic and orthotropic symmetries, respectively.
		Even-order tensors for these two symmetry classes are identical. For the fourth-order tensor $\mathbb{C}$, the addition of infinite rotation symmetry about the $z$-axis, enforces in-plane major symmetry ($C_{12} = C_{21}$), but out-of-plane odd elasticity survives ($C_{13} \neq C_{31}$). Accordingly, the stiffness tensor (containing 6 independent components) reads:
		\begin{equation}
			\mathbf{C}_\text{trans} = \begin{bmatrix}  
				C_{11} & C_{12} & C_{13} & 0 & 0 & 0\\
				C_{12} & C_{11} & C_{13} & 0 & 0 & 0\\ 
				C_{31} & C_{31} & C_{33} & 0 & 0 & 0\\ 
				0 & 0 & 0 & C_{44} & 0 & 0\\
				0 & 0 & 0 & 0 & C_{44} & 0\\ 
				0 & 0 & 0 & 0 & 0 & \frac{1}{2}(C_{11}-C_{12})
			\end{bmatrix} \, .
			\label{Cij-TH}
		\end{equation} 
		Similarly, the sixth-order stiffness tensor $\mathbb{A}$ for both transversely hemitropic and transversely isotropic cases share the same nonzero components as that of the orthorhombic case. However, the addition of the axis of infinite rotation symmetry imposes 53 internal identities that reduce their number of independent parameters to 31.
		These identities are categorized into three groups. 
		The first group comprises 30 direct transverse equalities reflecting the equivalence of the $x$ and $y$ axes:
		\begin{subequations}
			\label{PathA-TransverseH-A}
			\begin{align}
				A_{07\,07} &= A_{02\,02}, & A_{07\,08} &= A_{02\,01}, & A_{07\,09} &= A_{02\,03}, & A_{07\,16} &= A_{02\,17}, &A_{08\,07} &= A_{01\,02} \nonumber \\
				A_{08\,08} &= A_{01\,01}, & A_{08\,09} &= A_{01\,03}, & A_{08\,16} &= A_{01\,17}, &A_{09\,07} &= A_{03\,02}, & A_{09\,08} &= A_{03\,01} \nonumber \\
				A_{09\,09} &= A_{03\,03}, & A_{09\,16} &= A_{03\,17}, &A_{10\,05} &= A_{05\,10}, & A_{10\,10} &= A_{05\,05}, & A_{10\,13} &= A_{05\,14} \nonumber \\
				A_{10\,14} &= A_{05\,13}, &A_{14\,05} &= A_{13\,10}, & A_{14\,10} &= A_{13\,05}, & A_{14\,13} &= A_{13\,14}, & A_{14\,14} &= A_{13\,13} \nonumber \\
				A_{16\,07} &= A_{17\,02}, & A_{16\,08} &= A_{17\,01}, & A_{16\,09} &= A_{17\,03}, & A_{16\,16} &= A_{17\,17}, & A_{10\,15} &= A_{05\,15} \nonumber \\
				A_{14\,15} &= A_{13\,15}, & A_{15\,10} &= A_{15\,05}, & A_{15\,14} &= A_{15\,13}, & A_{11\,11} &= A_{04\,04}, & A_{08\,06} &= A_{01\,12}, 
				\label{PathA-TransverseH-A-1}
			\end{align}
			The second group comprises 13 out-of-plane isotropic shear relations:
			\begin{align}
				A_{09\,06} &= A_{03\,12} = \frac{1}{2}(A_{03\,01} - A_{03\,02}) \nonumber \\
				A_{06\,09} &= A_{12\,03} = A_{03\,12} + \frac{1}{2}(A_{01\,03} - A_{03\,01}) - \frac{1}{2}(A_{02\,03} - A_{03\,02}) \nonumber \\
				A_{17\,12} &= A_{16\,06} = \frac{1}{2}(A_{17\,01} - A_{17\,02}) \nonumber \\
				A_{06\,16} &= A_{12\,17} = A_{17\,12} + \frac{1}{2}(A_{01\,17} - A_{17\,01}) - \frac{1}{2}(A_{02\,17} - A_{17\,02}) \nonumber \\
				A_{18\,04} &= A_{18\,11} = \frac{1}{2}(A_{13\,05} - A_{13\,10}) \nonumber \\
				A_{04\,18} &= A_{11\,18} = A_{18\,11} + \frac{1}{2}(A_{05\,13} - A_{13\,05}) - \frac{1}{2}(A_{05\,14} - A_{14\,05}) \nonumber \\
				A_{18\,18} &= \frac{1}{2}(A_{13\,13} - A_{13\,14}).
				\label{PathA-TransverseH-A-2}
			\end{align}
			The third group comprises 10 planar cross-couplings and in-plane shear relations:
			\begin{align}
				A_{06\,08} &= A_{12\,01} = A_{01\,12} + \frac{1}{2}(A_{01\,02} - A_{02\,01}) \nonumber \\
				A_{07\,06} &= A_{02\,12} = \frac{1}{2}(A_{01\,01} - A_{02\,02}) - A_{01\,12} - \frac{1}{2}(A_{01\,02} - A_{02\,01}) \nonumber \\
				A_{06\,07} &= A_{12\,02} = A_{02\,12} + \frac{1}{2}(A_{01\,02} - A_{02\,01}) \nonumber \\
				A_{06\,06} &= A_{12\,12} = \frac{1}{4}(A_{01\,01} + A_{02\,02}) - \frac{1}{2}A_{01\,02} + \frac{1}{4}(A_{01\,02} - A_{02\,01}) \nonumber \\
				A_{04\,11} &= A_{11\,04} = A_{05\,05} - A_{05\,10} - A_{04\,04}.
				\label{PathA-TransverseH-A-3}
			\end{align}
		\end{subequations}

		Again, in the fifth-order stiffness tensor, the specific symmetry group dictates the nonzero components. In the chiral transversely hemitropic class ($\infty 22$), the infinite rotational symmetry imposes 16 internal constraints on the fifth-order tensor $\mathbb{F}$ (see Eq.~\eqref{PathA-TransverseH-F}), reducing it to 8 independent parameters: $\{F_{1\,04}, F_{2\,04}, F_{3\,04}, F_{4\,01}, F_{4\,02}, F_{4\,03}, F_{4\,17}, F_{6\,13}\}$. Incorporating these identities directly, the explicit matrix representation of $\mathbb{F}$ is given as:
		\begin{equation}
			\resizebox{\textwidth}{!}{
				$ \mathbf{F}_{\infty 22} = \begin{bmatrix}
					0 & 0 & 0 & F_{1\,04} & 0 & 0 & 0 & 0 & 0 & 0 & -F_{2\,04} & 0 & 0 & 0 & 0 & 0 & 0 & -F_{6\,13} \\
					0 & 0 & 0 & F_{2\,04} & 0 & 0 & 0 & 0 & 0 & 0 & -F_{1\,04} & 0 & 0 & 0 & 0 & 0 & 0 & F_{6\,13} \\
					0 & 0 & 0 & F_{3\,04} & 0 & 0 & 0 & 0 & 0 & 0 & -F_{3\,04} & 0 & 0 & 0 & 0 & 0 & 0 & 0 \\
					F_{4\,01} & F_{4\,02} & F_{4\,03} & 0 & 0 & 0 & 0 & 0 & 0 & 0 & 0 & (F_{4\,01}-F_{4\,02})/2 & 0 & 0 & 0 & 0 & F_{4\,17} & 0 \\
					0 & 0 & 0 & 0 & 0 & (F_{4\,02}-F_{4\,01})/2 & -F_{4\,02} & -F_{4\,01} & -F_{4\,03} & 0 & 0 & 0 & 0 & 0 & 0 & -F_{4\,17} & 0 & 0 \\
					0 & 0 & 0 & 0 & (F_{1\,04}-F_{2\,04})/2 & 0 & 0 & 0 & 0 & (F_{1\,04}-F_{2\,04})/2 & 0 & 0 & F_{6\,13} & -F_{6\,13} & 0 & 0 & 0 & 0
				\end{bmatrix} \, . $}
			\label{F-TH}
		\end{equation}
		For the achiral transversely isotropic class ($\infty/mmm$), the fifth-order has already vanished due to centrosymmetry
		\begin{equation}
			\mathbf{F}_{\infty/mmm} = \mathbf{0} \, .
			\label{F-TI}
		\end{equation}

		\subsection{Cubic Classes (Point Groups $23$ and $m\bar{3}$)}
		The cubic classes are generated by adding to the orthorhombic and orthotropic cases, a three-fold ($120^\circ$) rotation about the $x=y=z$ axis, which effectively enforces the cyclic permutations ($xyz\to yzx\to zxy$) to keep the stiffness tensors unchanged.
		Even-order tensors for these two symmetry groups are identical. For the fourth-order tensor $\mathbb{C}$ the addition of cubic symmetry preserves odd elasticity across the main diagonal ($C_{12} = C_{23} = C_{31} \neq C_{21} = C_{32} = C_{13}$) and is formulated via 4 independent parameters as
		\begin{equation}
			\mathbf{C}_{\text{cubic}} = \begin{bmatrix}  
				C_{11} & C_{12} & C_{21} & 0 & 0 & 0\\
				C_{21} & C_{11} & C_{12} & 0 & 0 & 0\\ 
				C_{12} & C_{21} & C_{11} & 0 & 0 & 0\\ 
				0 & 0 & 0 & C_{44} & 0 & 0\\
				0 & 0 & 0 & 0 & C_{44} & 0\\ 
				0 & 0 & 0 & 0 & 0 & C_{44}
			\end{bmatrix} \, .
			\label{Cij-cubic}
		\end{equation} 
		
		The sixth-order tensor $\mathbf{A}$ shares the nonzero components of the orthorhombic class. However, the cyclic permutations are shown to result in 56 equalities (listed in Eq.~\eqref{PathA-Cubic-A}). Imposing these equalities, the sixth-order stiffness tensor can be written using 28 independent components as:
		\setcounter{MaxMatrixCols}{18}
		\begin{equation}
			\resizebox{\textwidth}{!}{
				$ \mathbf{A}_{\text{cubic}} = \begin{bmatrix}
					A_{01\,01} & A_{01\,02} & A_{01\,03} & 0 & 0 & 0 & 0 & 0 & 0 & 0 & 0 & A_{01\,12} & 0 & 0 & 0 & 0 & A_{01\,17} & 0 \\
					A_{02\,01} & A_{02\,02} & A_{02\,03} & 0 & 0 & 0 & 0 & 0 & 0 & 0 & 0 & A_{02\,12} & 0 & 0 & 0 & 0 & A_{02\,17} & 0 \\
					A_{03\,01} & A_{03\,02} & A_{03\,03} & 0 & 0 & 0 & 0 & 0 & 0 & 0 & 0 & A_{03\,12} & 0 & 0 & 0 & 0 & A_{03\,17} & 0 \\
					0 & 0 & 0 & A_{04\,04} & 0 & 0 & 0 & 0 & 0 & 0 & A_{04\,11} & 0 & 0 & 0 & 0 & 0 & 0 & A_{04\,18} \\
					0 & 0 & 0 & 0 & A_{01\,01} & 0 & 0 & 0 & 0 & A_{01\,02} & 0 & 0 & A_{01\,03} & A_{01\,12} & A_{01\,17} & 0 & 0 & 0 \\
					0 & 0 & 0 & 0 & 0 & A_{01\,01} & A_{01\,02} & A_{01\,03} & A_{01\,12} & 0 & 0 & 0 & 0 & 0 & 0 & A_{01\,17} & 0 & 0 \\
					0 & 0 & 0 & 0 & 0 & A_{02\,01} & A_{02\,02} & A_{02\,03} & A_{02\,12} & 0 & 0 & 0 & 0 & 0 & 0 & A_{02\,17} & 0 & 0 \\
					0 & 0 & 0 & 0 & 0 & A_{03\,01} & A_{03\,02} & A_{03\,03} & A_{03\,12} & 0 & 0 & 0 & 0 & 0 & 0 & A_{03\,17} & 0 & 0 \\
					0 & 0 & 0 & 0 & 0 & A_{12\,01} & A_{12\,02} & A_{12\,03} & A_{12\,12} & 0 & 0 & 0 & 0 & 0 & 0 & A_{12\,17} & 0 & 0 \\
					0 & 0 & 0 & 0 & A_{02\,01} & 0 & 0 & 0 & 0 & A_{02\,02} & 0 & 0 & A_{02\,03} & A_{02\,12} & A_{02\,17} & 0 & 0 & 0 \\
					0 & 0 & 0 & A_{11\,04} & 0 & 0 & 0 & 0 & 0 & 0 & A_{04\,04} & 0 & 0 & 0 & 0 & 0 & 0 & A_{04\,11} \\
					A_{12\,01} & A_{12\,02} & A_{12\,03} & 0 & 0 & 0 & 0 & 0 & 0 & 0 & 0 & A_{12\,12} & 0 & 0 & 0 & 0 & A_{12\,17} & 0 \\
					0 & 0 & 0 & 0 & A_{03\,01} & 0 & 0 & 0 & 0 & A_{03\,02} & 0 & 0 & A_{03\,03} & A_{03\,12} & A_{03\,17} & 0 & 0 & 0 \\
					0 & 0 & 0 & 0 & A_{12\,01} & 0 & 0 & 0 & 0 & A_{12\,02} & 0 & 0 & A_{12\,03} & A_{12\,12} & A_{12\,17} & 0 & 0 & 0 \\
					0 & 0 & 0 & 0 & A_{17\,01} & 0 & 0 & 0 & 0 & A_{17\,02} & 0 & 0 & A_{17\,03} & A_{17\,12} & A_{17\,17} & 0 & 0 & 0 \\
					0 & 0 & 0 & 0 & 0 & A_{17\,01} & A_{17\,02} & A_{17\,03} & A_{17\,12} & 0 & 0 & 0 & 0 & 0 & 0 & A_{17\,17} & 0 & 0 \\
					A_{17\,01} & A_{17\,02} & A_{17\,03} & 0 & 0 & 0 & 0 & 0 & 0 & 0 & 0 & A_{17\,12} & 0 & 0 & 0 & 0 & A_{17\,17} & 0 \\
					0 & 0 & 0 & A_{04\,11} & 0 & 0 & 0 & 0 & 0 & 0 & A_{11\,04} & 0 & 0 & 0 & 0 & 0 & 0 & A_{04\,04}
				\end{bmatrix} \, . $}
			\label{A-cubic}
		\end{equation}
		
		The fifth-order tensor is nonzero only in the chiral class, where the cubic symmetry imposes 16 equalities between the components of $\mathbb{F}$ (see Eq.~\eqref{PathA-Cubic-F}), reducing its independent components to the following 8 parameters: $\{F_{1\,04}, F_{1\,11}, F_{2\,04}, F_{4\,01}, F_{4\,02}, F_{4\,03}, F_{4\,12}, F_{4\,17}\}$. Incorporating these identities directly, the explicit matrix representation is given by:
		\begin{equation}
			\resizebox{\textwidth}{!}{
				$ \mathbf{F}_{23} = \begin{bmatrix}
					0 & 0 & 0 & F_{1\,04} & 0 & 0 & 0 & 0 & 0 & 0 & F_{1\,11} & 0 & 0 & 0 & 0 & 0 & 0 & F_{2\,04} \\
					0 & 0 & 0 & F_{2\,04} & 0 & 0 & 0 & 0 & 0 & 0 & F_{1\,04} & 0 & 0 & 0 & 0 & 0 & 0 & F_{1\,11} \\
					0 & 0 & 0 & F_{1\,11} & 0 & 0 & 0 & 0 & 0 & 0 & F_{2\,04} & 0 & 0 & 0 & 0 & 0 & 0 & F_{1\,04} \\
					F_{4\,01} & F_{4\,02} & F_{4\,03} & 0 & 0 & 0 & 0 & 0 & 0 & 0 & 0 & F_{4\,12} & 0 & 0 & 0 & 0 & F_{4\,17} & 0 \\
					0 & 0 & 0 & 0 & 0 & F_{4\,01} & F_{4\,02} & F_{4\,03} & F_{4\,12} & 0 & 0 & 0 & 0 & 0 & 0 & F_{4\,17} & 0 & 0 \\
					0 & 0 & 0 & 0 & F_{4\,01} & 0 & 0 & 0 & 0 & F_{4\,02} & 0 & 0 & F_{4\,03} & F_{4\,12} & F_{4\,17} & 0 & 0 & 0
				\end{bmatrix} \, . $}
			\label{F-cubic}
		\end{equation}
		For the achiral cubic class ($m\bar{3}$), the presence of a center of inversion dictates that
		\begin{equation}
			\mathbf{F}_{m\bar{3}} = \mathbf{0} \, .
			\label{F-achiralcubic}
		\end{equation}

		\subsection{Hemitropic ($SO(3)$) and Isotropic ($O(3)$)}
		For continuous 3D rotational symmetry, the fourth-order tensor $\mathbf{C}$ (2 parameters) is unconditionally major-symmetric and can be represented using Lam\'e constants as
		\begin{equation}
			\mathbf{C}_{\text{iso}} = \begin{bmatrix}  
				c_1+2c_2 & c_1 & c_1 & 0 & 0 & 0\\
				c_1 & c_1+2c_2 & c_1 & 0 & 0 & 0\\ 
				c_1 & c_1 & c_1+2c_2 & 0 & 0 & 0\\ 
				0 & 0 & 0 & c_2 & 0 & 0\\
				0 & 0 & 0 & 0 & c_2 & 0\\ 
				0 & 0 & 0 & 0 & 0 & c_2
			\end{bmatrix} \, .
			\label{Cij-iso}
		\end{equation} 
		Similarly, the sixth-order tensor $\mathbf{A}$ is identical for both hemitropic and isotropic classes. It shares the nonzero locations of the orthorhombic class and its components follow the linear combinations of Kronecker deltas shown in Eq.~\eqref{PathA-Hemi-A}. It is noted that in both cases, odd elasticity is preserved via the split of the first stiffness coefficient into $a_{1a} \neq a_{1b}$.

		For the chiral hemitropic class ($SO(3)$), the fifth-order tensor $\mathbb{F}$ is populated by the single chiral modulus $f_1$ as:
		\begin{equation}
			\resizebox{\textwidth}{!}{
				$ \mathbf{F}_{SO(3)} = 2f_1 \begin{bmatrix}
					0 & 0 & 0 & 1 & 0 & 0 & 0 & 0 & 0 & 0 & -1 & 0 & 0 & 0 & 0 & 0 & 0 & -1 \\
					0 & 0 & 0 & 1 & 0 & 0 & 0 & 0 & 0 & 0 & -1 & 0 & 0 & 0 & 0 & 0 & 0 & 1 \\
					0 & 0 & 0 & 1 & 0 & 0 & 0 & 0 & 0 & 0 & -1 & 0 & 0 & 0 & 0 & 0 & 0 & 0 \\
					1 & 1 & 1 & 0 & 0 & 0 & 0 & 0 & 0 & 0 & 0 & 0 & 0 & 0 & 0 & 0 & 1 & 0 \\
					0 & 0 & 0 & 0 & 0 & 0 & -1 & -1 & -1 & 0 & 0 & 0 & 0 & 0 & 0 & -1 & 0 & 0 \\
					0 & 0 & 0 & 0 & 0 & 0 & 0 & 0 & 0 & 0 & 0 & 0 & 1 & -1 & 0 & 0 & 0 & 0
				\end{bmatrix} \, . $}
			\label{F-SO3}
		\end{equation}
		For the fully isotropic achiral class ($O(3)$), the fifth-order stiffness tensors vanish:
		\begin{equation}
			\mathbf{F}_{O(3)} = \mathbf{0} \, .
			\label{F-iso}
		\end{equation}

		\subsection{Expanded Symmetry Classification}
		\label{app:expanded_classification}
		
		The main text details the explicit matrix representations for a representative hierarchy of elastic symmetries. In standard crystallography, there are 32 discrete point groups. In classical Cauchy elasticity, the fourth-order stiffness tensor ($\mathbb{C}$) collapses these 32 groups into 8 distinct elastic symmetry classes \cite{forte1996symmetry}, due to the centrosymmetry and the fourth order of the stiffness tensor. As an example, in fourth-order tensors, $60^\circ$ rotation symmetry (hexagonal symmetry) and continuous rotational symmetry (transversely isotropic symmetry) are identical).
		
		In strain-gradient elasticity, however, the higher order of the stiffness tensors and the odd order of the fifth-order coupling tensors break this degeneracy and imply different conditions for different symmetry point groups. It has been shown by Auffray et al. \cite{auffray2019complete} that when the symmetries of the complete set of stiffness tensors of second gradient elasticity $(\mathbb{C}, \mathbb{F}$ or $\mathbb{G}$, and $\mathbb{A})$ are combined, the mathematical space of strain-gradient elasticity extends to 48 distinct symmetry classes. Table \ref{tab:Expanded_Summary} provides a comprehensive list of the number of independent components of strain-gradient odd elasticity tensors for the fundamental elastic symmetry groups.

		\begin{table}[htpb]
			\centering
			\renewcommand{\arraystretch}{1.4}
			\resizebox{\textwidth}{!}{%
				\begin{tabular}{l | c | c | c | c | c | c | c | c | c | c | c | c | c}
					\hline
					\multicolumn{14}{c}{\textbf{Chiral branch}} \\
					\hline
					Crystal System & Triclinic & Monoclinic & Orthorhombic & \multicolumn{2}{c|}{Tetragonal} & \multicolumn{2}{c|}{Trigonal} & \multicolumn{2}{c|}{Hexagonal} & Transverse & \multicolumn{2}{c|}{Cubic} & Isotropic \\
					\hline
					Point Group & $1$ & $2$ & $222$ & $4$ & $422$ & $3$ & $32$ & $6$ & $622$ & $\infty 22$ & $23$ & $432$ & $SO(3)$ \\
					\hline
					$\mathbb{C}$ (Odd) & 36 & 20 & 12 & 10 & 7 & 12 & 8 & 8 & 6 & 6 & 4 & 3 & 2 \\
					$\mathbb{C}$ (Even) & 21 & 13 & 9 & 7 & 6 & 7 & 6 & 5 & 5 & 5 & 3 & 3 & 2 \\
					$\mathbb{F}, \mathbb{G}$ (Odd) & 108 & 52 & 24 & 26 & 11 & 36 & 16 & 20 & 8 & 8 & 8 & 3 & 1 \\
					$\mathbb{A}$ (Odd) & 324 & 164 & 84 & 82 & 43 & 108 & 56 & 60 & 32 & 31 & 28 & 15 & 6 \\
					$\mathbb{A}$ (Even) & 171 & 91 & 51 & 45 & 28 & 57 & 34 & 33 & 22 & 21 & 17 & 11 & 5 \\
					\hline
					\hline
					\multicolumn{14}{c}{\textbf{Achiral branch}} \\
					\hline
					Crystal System & Monoclinic & Orthorhombic & Orthotropic & \multicolumn{2}{c|}{Tetragonal} & \multicolumn{2}{c|}{Trigonal} & \multicolumn{2}{c|}{Hexagonal} & Transverse & \multicolumn{2}{c|}{Cubic} & Isotropic \\
					\hline
					Point Group & $m$ & $mm2$ & $mmm$ & $4/m$ & $4/mmm$ & $\bar{3}$ & $\bar{3}m$ & $6/m$ & $6/mmm$ & $\infty/mmm$ & $m\bar{3}$ & $m\bar{3}m$ & $O(3)$ \\
					\hline
					$\mathbb{C}$ (Odd) & 20 & 12 & 12 & 10 & 7 & 12 & 8 & 8 & 6 & 6 & 4 & 3 & 2 \\
					$\mathbb{C}$ (Even) & 13 & 9 & 9 & 7 & 6 & 7 & 6 & 5 & 5 & 5 & 3 & 3 & 2 \\
					$\mathbb{F}, \mathbb{G}$ (Odd) & 56 & 28 & 0 & 0 & 0 & 0 & 0 & 0 & 0 & 0 & 0 & 0 & 0 \\
					$\mathbb{A}$ (Odd) & 164 & 84 & 84 & 82 & 43 & 108 & 56 & 60 & 32 & 31 & 28 & 15 & 6 \\
					$\mathbb{A}$ (Even) & 91 & 51 & 51 & 45 & 28 & 57 & 34 & 33 & 22 & 21 & 17 & 11 & 5 \\
					\hline
				\end{tabular}%
			}
			\caption{Number of independent components for the strain-gradient odd elasticity tensors across the fundamental crystallographic and continuous point groups. Values are presented for both the nonconservative odd elastic framework and the classical conservative (even) limit. For fifth-order tensors, the values represent the parameters for either $\mathbb{F}$ or $\mathbb{G}$ independently.}
			\label{tab:Expanded_Summary}
		\end{table}

		The table contains the elastic symmetries studied in the main text along with the following additional elastic symmetries:
		\begin{itemize}
			\item \textbf{Tetragonal:} Generated by a 4-fold ($90^\circ$) rotation about the $z$-axis. Point group $4$ contains only this rotation, while $422$ adds an orthogonal 2-fold ($180^\circ$) rotation about the $x$-axis.
			\item \textbf{Trigonal:} Generated by a 3-fold ($120^\circ$) rotation about the $z$-axis. Point group $3$ contains only this rotation, while $32$ adds an orthogonal 2-fold ($180^\circ$) rotation about the $x$-axis.
			\item \textbf{Hexagonal:} Generated by a 6-fold ($60^\circ$) rotation about the $z$-axis. Point group $6$ contains only this rotation, while $622$ adds an orthogonal 2-fold ($180^\circ$) rotation about the $x$-axis.
			\item \textbf{Cubic:} Point group $23$ is derived by adding to the orthorhombic ($222$) symmetry, a 3-fold ($120^\circ$) rotation about the $x=y=z$ axis. Point group $432$ upgrades the orthogonal 2-fold rotations of the underlying orthorhombic class to 4-fold ($90^\circ$) rotations.
		\end{itemize}

		The achiral counterparts of these symmetry point groups are constructed by superimposing centrosymmetry onto their respective chiral counterparts. 
		The results presented in Table \ref{tab:Expanded_Summary} are categorized into the chiral branch (rotations only) and the achiral branch (involving reflection symmetries). Similar to Table \ref{tab:Master_Summary}, for each symmetry class, we also include the number of independent components for conservative (even) elasticity.

		\section{Sufficient Conditions for the Hermiticity of the Acoustic Tensor}
		\label{app:hermiticity}
		
		To guarantees conservative wave propagation in the material, the acoustic tensor needs to be Hermitian, resulting in purely real eigenvalues ($\omega^2 \in \mathbb{R}$). For the tensor $\mathbf{\Gamma}$ to be Hermitian, it must be equal to its conjugate transpose ($\Gamma_{ik} = \bar{\Gamma}_{ki}$), which in turn requires the real part to be symmetric ($\Gamma_{ik}^\text{R} = \Gamma_{ki}^\text{R}$) and the imaginary part to be antisymmetric ($\Gamma_{ik}^\text{I} = -\Gamma_{ki}^\text{I}$). Here, we demonstrate two independently sufficient conditions, each guaranteeing Hermiticity of $\mathbf{\Gamma}$.
		
		\subsection{Condition 1: Classical Conservative Reciprocity}
		In classical conservative elasticity, the existence of a strain energy density potential enforces major symmetry on the even-order stiffness tensors (i.e. $C_{ijkl} = C_{klij}$ and $A_{ijklmn} = A_{lmnijk}$) and reciprocity of the fifth-order stiffness tensors (i.e. $G_{ijklm} = F_{lmijk}$). 
		
		Accounting for the major symmetry of the even-order stiffness tensors yields:
		\begin{equation}
			\begin{aligned}
				\Gamma_{ki}^\text{R} &= C_{kjil} k_j k_l + A_{kjlimn} k_j k_l k_m k_n \\
				& = C_{klij} k_j k_l + A_{kmnijl} k_j k_l k_m k_n\\
				& = C_{ijkl} k_j k_l + A_{ijlkmn} k_j k_l k_m k_n = \Gamma_{ik}^\text{R}\, ,
			\end{aligned}
		\end{equation}
		signifying the symmetry of the real part of $\mathbf{\Gamma}$.
		Similarly, accounting for the reciprocity of the fifth-order stiffness tensors yields:
		\begin{equation}
			\begin{aligned}
				\Gamma_{ki}^\text{I} &= (F_{kjilm} - G_{kjlim}) k_j k_l k_m \\
				& = (G_{ilmkj} - F_{imkjl}) k_j k_l k_m\\
				& = (G_{ijlkm} - F_{ijklm}) k_j k_l k_m = - \Gamma_{ik}^\text{I} \, ,
			\end{aligned}
		\end{equation}
		signifying the antisymmetry of the imaginary part of $\mathbf{\Gamma}$.
		Accordingly, provided the material is conservative, the acoustic tensor is unconditionally Hermitian.
		
		\subsection{Condition 2: Hemitropic Elastic Symmetry}
		Even if thermodynamic reciprocity is revoked (odd elasticity), the acoustic tensor remains Hermitian if the material possesses continuous 3D rotational symmetry (either complete isotropy $O(3)$ or chial isotropy (hemitropy) $SO(3)$). 
		
		Substituting the relationships for the stiffness tensors of hemitropic materials, Eq.~\eqref{PathA-Hemi-C}-\eqref{PathA-Hemi-G}, into Eq.~\eqref{Eq-Acoustic-Tensor} demonstrates that: (i) The real part of the acoustic tensor, $\Gamma_{ik}^\text{R}$, which contains even-order stiffness tensors multiplied an even number of times with $\mathbf{k}$, simplifies to a linear combination of $\delta_{ik}$ and $k_i k_k$. Since both of these represent symmetric tensors, $\Gamma_{ik}^\text{R}$ is symmetric. (ii) The imaginary part of the acoustic tensor, $\Gamma_{ik}^\text{I}$, which contains odd-order stiffness tensors multiplied three times with $\mathbf{k}$, simplifies to a linear combination of $\epsilon_{ikj} k_j$, which is antisymmetric. Accordingly, even in nonconservative odd elastic materials, provided that the material has 3D continuous rotational symmetry, its acoustic tensor remains Hermitian, and will guarantee conservative wave propagation.

		\section{Transversely Hemitropic Material Parameters}
		\label{app:THparams}

		In a transversely hemitropic material where the $z$-axis is the axis of symmetry, the fourth-order tensor $\mathbb{C}$ is defined using the following 6 independent components in Voigt notation $\{C_{11}, C_{33}, C_{12}, C_{13}, C_{31}, C_{44}\}$. To derive the stiffness tensors of the transversely hemitropic material for the numerical example, we use the hemitropic material of Section \ref{sec:res_hemitropic} as the baseline. We then kept the in-plane components identical to the hemitropic case ($C_{11} =  c_1 + 2c_2$, $C_{12} = c_1$), but stiffened the out-of-plane normal and shear components by $20\%$ ($C_{33} = 1.2 C_{11}$, $C_{44} = 1.2 c_2$). For the transverse cross-couplings, in the passive conservative limit, we stiffen the component corresponding to the axial direction as $C_{13} = C_{31} = 2c_1$. To introduce odd elasticity, we perturb one of these conjugate components to be three times larger, setting $C_{13} = 6 c_1$ while keeping $C_{31} = 2 c_1$. 
		
		Further, transversely hemitropic symmetry dictates 8 independent parameters for each of the fifth-order tensors, $\mathbb{F}$ and $\mathbb{G}$. In the hemitropic baseline, Eqs.~\eqref{PathA-Hemi-F} and \eqref{PathA-Hemi-G} show that all 8 parameters reduce to $2f_1$ and $2g_1$, respectively. To break this hemitropy and establish transverse hemitropy, it is sufficient to perturb the components associated with the direction of the axis of rotational symmetry, i.e., the $z$-axis. Accordingly, we have stiffened these parameters by $20\%$, setting $F_{3\,04} = F_{4\,03} = 2.4f_1$, and $G_{04\,3} = G_{03\,4} = 2.4g_1$. The remaining six independent components of both fifth-order stiffness tensors are kept unchanged: $F_{1\,04} = F_{2\,04} = F_{4\,01} = F_{4\,02} = F_{4\,17} = F_{6\,13} = 2f_1$ and $G_{04\,1} = G_{04\,2} = G_{01\,4} = G_{02\,4} = G_{17\,4} = G_{13\,6} = 2g_1$.

		The sixth-order stiffness tensor $\mathbb{A}$ for transversely hemitropic odd elasticity has $84$ nonzero components, out of which 31 parameters are independent.
		By collecting the 84 nonzero components of $\mathbb{A}$ in an array, $\mathbf{y}$, the 53 transversely-hemitropic linear constraints, Eq.~\eqref{PathA-TransverseH-A}, can be formulated using a homogeneous linear system of equations, $\mathbf{M}\mathbf{y} = \mathbf{0}$, where $\mathbf{M}$ is a $53 \times 84$ coefficients matrix. The $84 \times 31$ basis matrix, $\mathbf{N}$, which linearly maps the $31 \times 1$ array of \textit{independent} transversely hemitropic parameters, $\mathbf{c}$, onto the full list of components as $\mathbf{y} = \mathbf{N}\mathbf{c}$, is then formulated using the null-space projection method \cite{zheng1994theory,browaeys2004decomposition}.
		
		To enforce nonconservative behavior on the baseline conservative material, similar to the isotropic case in Section \ref{sec:res_iso}, we break Maxwell-Betti reciprocity by enforcing $a_{1a} = a_{1}$ and $a_{1b} = 1.5 a_{1}$. We then generate the full list of 84 nonzero components of the baseline isotropic odd elastic sixth-order stiffness tensor, $\mathbf{y}_{\text{iso}}$. By solving the system $\mathbf{y}_{\text{iso}} = \mathbf{N}\mathbf{c}_{\text{iso}}$, we identify the 31 independent parameters, $\mathbf{c}_{\text{iso}}$, that correspond to this isotropic state. 
		To break the isotropy of $\mathbf{c}_{\text{iso}}$, we perturb all its components to derive the transversely hemitropic stiffness components $\mathbf{c}$ as $\mathbf{c}_{\text{TH}} = \mathbf{c}_{\text{iso}} + \varepsilon ||\mathbf{c}_{\text{iso}}|| \mathrm{sgn}(\mathbf{c}_{\text{iso}})$, ensuring that $\mathbf{c}_{\text{TH}}$ does not inherit the symmetries of the isotropic baseline ($\varepsilon$ denotes the perturbation value, for which we have used $0.03$ here). The full set of the nonzero components of $\mathbb{A}$ for the transversely hemitropic odd elastic material is then derived as $\mathbf{y}_{\text{TH}} = \mathbf{N}\mathbf{c}_{\text{TH}}$.

	\end{appendices}
	
	
	\bibliographystyle{RS} 
	\bibliography{ReferencesList}

\end{document}